\newcommand{\be}{\begin{equation}}
\newcommand{\ee}{\end{equation}}
\newcommand{\bea}{\begin{eqnarray}}
\newcommand{\eea}{\end{eqnarray}}

\newcommand{\gsim}{\lower.7ex\hbox{$\;\stackrel{\textstyle>}{\sim}\;$}}
\newcommand{\lsim}{\lower.7ex\hbox{$\;\stackrel{\textstyle<}{\sim}\;$}}

\newcommand{\mrm}{\mathrm}
\newcommand{\df}{\dfrac}

\documentclass[aps,prev,twocolumn,preprintnumbers,floatfix,nofootinbib]{revtex4-1}
\usepackage[pdftex]{graphicx}
\usepackage{mathrsfs}
\usepackage{amssymb}
\usepackage{verbatim}
\usepackage{color}
\usepackage{mathrsfs}
\usepackage{amsmath, amsthm, amssymb}
\usepackage{hyperref}

\def\stacksymbols #1#2#3#4{\def\theguybelow{#2}
    \def\vp{\lower#3pt}
    \def\sp{\baselineskip0pt\lineskip#4pt}
    \mathrel{\mathpalette\intermediary#1}}

\def\intermediary#1#2{\vp\vbox{\sp
     \everycr={}\tabskip0pt
     \halign{$\mathsurround0pt#1\hfil##\hfil$\crcr#2\crcr
              \theguybelow\crcr}}}

\def\be{\begin{equation}}
\def\ee{\end{equation}}
\def\bea{\begin{eqnarray}}
\def\eea{\end{eqnarray}}

\def\sp{\;\;\;,\;\;\;}

\def\mrm{\mathrm}

\def\lsim{\raise0.3ex\hbox{$\;<$\kern-0.75em\raise-1.1ex\hbox{$\sim\;$}}}
\def\gsim{\raise0.3ex\hbox{$\;>$\kern-0.75em\raise-1.1ex\hbox{$\sim\;$}}}

\def\inbar{\,\vrule height1.5ex width.4pt depth0pt}

\def\IC{\relax\hbox{$\inbar\kern-.3em{\rm C}$}}
\def\IQ{\relax\hbox{$\inbar\kern-.3em{\rm Q}$}}
\def\IR{\relax{\rm I\kern-.18em R}}
 \font\cmss=cmss10 \font\cmsss=cmss10 at 7pt
\def\IZ{\relax\ifmmode\mathchoice
 {\hbox{\cmss Z\kern-.4em Z}}{\hbox{\cmss Z\kern-.4em Z}}
 {\lower.9pt\hbox{\cmsss Z\kern-.4em Z}}
 {\lower1.2pt\hbox{\cmsss Z\kern-.4em Z}}\else{\cmss Z\kern-.4em Z}\fi}

\def\comment#1{}

\def\u1x{U(1)_X}
\newcommand{\nc}{\newcommand}
\nc{\LL}{L}
\nc{\vv}{\tilde{v}}
\nc{\ccdot}{\!\cdot\!}
\nc{\gsm}{G_{SM}}
\nc{\vfive}{\mathbf{5}\oplus\mathbf{\overline{5}}}
\nc{\vten}{\mathbf{10}\oplus\mathbf{\overline{10}}}
\nc{\zhol}{Z^{\rm hol}}
\nc{\xfb}{\,{\rm fb}}

\newcommand{\lsi}{\,\raisebox{-0.13cm}{$\stackrel{\textstyle<}
{\textstyle\sim}$}\,}
\newcommand{\gsi}{\,\raisebox{-0.13cm}{$\stackrel{\textstyle> 
}
{\textstyle\sim}$}\,}

\begin{document}

%
%

\preprint{LPT--Orsay-12-40 }
\preprint{IFT-UAM/CSIC-12-47}
\preprint{ULB/TH-12-10}
\preprint{SACLAY-T12/044}

\title{Complementarity of Galactic radio and collider data in constraining WIMP dark matter models}
\author{Yann Mambrini$^{a}$}
\email{yann.mambrini@th.u-psud.fr}
\author{Michel H.G. Tytgat$^{b}$}
\email{mtytgat@ulb.ac.be}
\author{Gabrijela Zaharijas$^{c,d}$}
\email{gzaharij@ictp.it}
\author{Bryan Zald\'\i var$^{e}$}
\email{bryan.zaldivar@uam.es}

\vspace{1cm}
\affiliation{
${}^a$ Laboratoire de Physique Th\'eorique 
Universit\'e Paris-Sud, F-91405 Orsay, France \\
${}^b$ Service de Physique Th\'eorique, Universit\'e Libre de Bruxelles, CP225, Bld du Triomphe, 1050 Brussels, Belgium\\
${}^c$ The Abdus Salam International Centre for Theoretical Physics, Strada Costiera 11, I-34014 Trieste, Italy \\
${}^d$ Institut de Physique Th\'eorique, CEA/Saclay, F-91191 Gif sur Yvette, France \\
${}^e$Instituto de Fisica Teorica, IFT-UAM/CSIC, 
Nicolas Cabrera 15, UAM
Cantoblanco, 28049 Madrid, Spain 
}

\begin{abstract}
{In this work we confront  dark matter models to constraints that may be derived from radio synchrotron radiation from the Galaxy, taking into account the astrophysical uncertainties and we compare these to bounds set by {accelerator and complementary indirect dark matter searches}
. Specifically we apply our analysis to three popular particle physics models. First, a generic effective operator approach, in which case we set bounds on the corresponding mass scale, and then, two specific UV completions, the $Z^\prime$ and Higgs portals. We show that for many candidates, the radio synchrotron limits are competitive with the other searches, and could even give the strongest constraints (as of today) with some reasonable assumptions regarding the astrophysical uncertainties.} 
\end{abstract}

\maketitle

\section{Introduction}
\label{sec:intro}

Dark Matter (DM)  is one of the most important issues in particle physics and cosmology and understanding its nature will likely play an essential r\^ole in our comprehension of both fundamental interactions and the structure of the Universe.
Over the years, a Weakly Interacting Massive Particle (WIMP) has emerged as one of the favourite candidates,  in part due to the natural explanation of its cosmological abundance through thermal freeze-out (see e.g. \cite{Bergstrom:2000pn,Bertone:2004pz,Bertone:2010zz}). 
While, as of today, we have no other evidence for DM than through its gravitational manifestations,
the alleged weak interaction of WIMPs have open the possibility to actually observe DM experimentally.
Several strategies have been proposed, and are  actively pursued, to search for WIMPs. Direct detection experiments, such as CDMS \cite{Ahmed:2010wy} and  XENON100 \cite{Moon:2010wq}, are dedicated to the search of DM in the vicinity of the solar system. These are supplemented by multi-purpose particle physics experiments at colliders, most notably the LHC, where DM is expected to be produced, and to manifest itself as missing energy, in collisions. In particular, analysis of single-photon or mono-jet events with missing energy  have recently proved to give very pertinent constraints on WIMP mass and interactions \cite{Goodman:2010ku,Bai:2010hh,Fox:2011fx,Fox:2011pm,Chatrchyan:2012te}. A radically different, and a complementary approach is to search for indirect detection of WIMPs, through the remnants of their annihilation (or decays) in astrophysical environments, like the Galactic Centre (GC) of the Milky Way, nearby dwarf spheroidal galaxies (dSphs) or in general any dense region of the Universe. The possible remnants, or messengers,  are high energy neutrinos, anti-matter in cosmic rays (CR), and gamma-rays, or more generally an injection of energy of charged particles in the early universe. 



In \cite{Mambrini:2011pw} some of us analysed  bounds on effective couplings between dark matter and SM particles
from single-photon and mono-jet signals, at LEP and the LHC respectively. Such studies are usually compared to exclusion limits set by direct detection experiments (see also {\em e.g.} \cite{Bai:2010hh,Fox:2011pm,Chatrchyan:2012te}). In general terms, the colliders data are comparatively more constraining for low mass dark matter candidates, in particular below the threshold of direct detection experiments, while the latter are more constraining at higher masses, where dark matter production at colliders is impeded. Interestingly, indirect searches tend also to be most constraining for low mass dark matter candidates. This is essentially because the flux of particles produced by dark matter annihilation (gamma-rays, etc.) is proportional to the inverse DM mass to the square. In particular, interesting constraints have been set on the annihilation cross section of DM based on the measured synchrotron radiation from the inner regions of the Milky Way. {While synchrotron radiation constraints on DM have already been much studied in the literature (e.g. in \cite{Bergstrom:2008ag,Bringmann:2009ca,Borriello:2008gy,Boehm:2010kg,Fornengo:2011iq}), to our knowledge no analysis of specific particle physics models implications have been made so far}. Concretely in the present work we confront the constraints from synchrotron radiation in the Galaxy at radio frequencies to those set by colliders data. In particular we study to which extend they are complementary. Our analysis is based both on an effective operator approach (so we put limits on energy scales) and on two specific DM models, the so-called $Z^\prime$ and Higgs portals. We also show how colliders and radio synchrotron radiation limits compare to bounds set by {\it Fermi} LAT based on dSphs \cite{Ackermann:2011wa}, and to constraints imposed on DM annihilations from the effect on the CMB anisotropies \cite{Galli:2011rz}.


The constraints from colliders and those from indirect searches do not quite stand on the same footing. In particular, although the radio data are potentially strongly constraining for rather light DM candidates -- as we show also in this work -- the modelling of DM induced radio fluxes suffers from several sources of astrophysical uncertainties. 
We explore those by using  both a semi-analytic approach, which allows to control, for instance, the dependence of the radio flux on the magnitude of the magnetic fields in our region of interest, and a full numerical calculation as implemented in the \texttt{GALPROP}\footnote{http://galprop.stanford.edu/webrun.php} code \cite{Porter:2008ve,Vladimirov:2010aq} which allows  to explore the full set of CR propagation parameters and up-to-date energy losses, thus cross checking the semi-analytic method and calibrating its parameters. To gauge the impact of CR propagation parameters on synchrotron signals, we sample a range of CR propagation parameters sets using both those derived to probe uncertainty in the local CR fluxes, and used traditionally to bracket this type of uncertainty (MIN, MED, MAX set of parameters \cite{Donato:2003xg}), and CR propagation sets which were recently shown to give a good description of the gamma-ray \cite{FermiLAT:2012aa,Trotta:2010mx} and radio \cite{Strong:2011wd} data and which are therefore suited to test electron propagation parameters in the inner regions of our galaxy. 

We present our results as a function of the systematic uncertainties on the
modelling of the backgrounds due to standard astrophysics, illustrating potential future improvements of the limits based on radio data. 

The paper is organised as follows. In section \ref{sec:synchrotron} we review the formalism of the semi-analytical approach used in this work, and also present the setup for the full numerical study.  In Section \ref{sec:astro} we discuss at lengths the astrophysical framework. In part \ref{sec:astroU} we discuss the astrophysical uncertainties, while in part \ref{sec:cross} we discuss our cross-check between semi-analytical and numerical approaches. In part \ref{sec:DMprofiles} we show the dependence of the synchrotron flux with different choices of Dark Matter profile, and Part \ref{sec:CRpars} is devoted to study, numerically, the impact of different choices of CR diffusion and propagation parameters to the synchrotron signal. Part \ref{sec:magn} is a semi-analytical (cross-checked by numerical) study of the dependence of the synchrotron flux on the magnetic field normalisation. Then, 
section \ref{sec:models} is devoted to the discussion of synchrotron constraints on particle physics models.  In part \ref{sec:effectiveOperators} we adopt a generic, also known as model-independent, approach based on effective operators. There we confront the synchrotron constraints we derive with other indirect searches and colliders constraints.  In parts  \ref{sec:HiggsPortal} and \ref{sec:ZpPortal}  we consider two specific UV completions,  the Higgs and Z$^\prime$ portal,  to which we apply the synchrotron constraints. While the conclusions bear some resemblance, the constraints are quite distinct for the two models. Finally we give our conclusion and prospects in \ref{sec:conclusion}.

\section{Synchrotron radio emission in the Milky Way}
\label{sec:synchrotron}

In the very high frequency radio (VHF) to microwave frequency range ({\em i.e.} $10\,$MHz $-$ 300\,GHz), various astrophysical processes contribute to the observed diffuse emission. The radio sky at frequencies below $\sim 20$ GHz is dominated by synchrotron emission of CR electrons (accelerated in e.g. supernovae shocks) on Galactic magnetic fields. However thermal bremsstrahlung (free-free emission) of electrons on the galactic ionised gas also contributes in this range. At higher frequencies the Cosmic Microwave Background (CMB) and its anisotropies represent the main signal. Emission by small grains of vibrating or spinning dust becomes relevant at even higher frequencies, starting approximately at $\gsi 60$ GHz. 

Synchrotron signal of electron by-products of dark matter WIMP self-annihilation is generally expected to fall in the 100 MHz $-$100 GHz range (see for example \cite{Borriello:2008gy}). However, it has been noted in \cite{Boehm:2010kg,Fornengo:2011iq} that  for light dark matter candidates it is beneficial to consider lower frequencies. For instance, for a magnetic field of $\mu$G strength, as typical for the Galaxy, electrons with energies $\lsi$ 10 GeV generate synchrotron emission which peaks at tens of MHz (see  Figure 4 in \cite{Fornengo:2011iq}  or Eq.~(\ref{Ecrit}) below),  and following \cite{Fornengo:2011iq} we focus on a 45 MHz survey \cite{Guzman:2010da}.

Additional motivation to consider the 45 MHz survey is that it covers a large region of the sky (for a visual representation of the coverage of this and other radio surveys, see \cite{deOliveiraCosta:2008pb}). In the past, observations of the Galactic Centre compact radio source (SgrA*) with high angular resolution surveys have been used  to constrain a possible DM signal (but at higher frequencies, for example a 408 MHz survey \cite{Haslam:1982zz} was used in 
\cite{Bergstrom:2008ag,Bringmann:2009ca} and a 330 MHz observation of the SgrA* \cite{Nord:2004mi} in \cite{Boehm:2010kg}). 
Indeed the synchrotron flux is expected to be strongly enhanced close to the Galactic centre, as 
cusps in a DM density are generically found in DM simulations in central regions of DM halos,
and therefore strong constraints on DM annihilation cross section may be placed by considering radio emission from SgrA*. However, the mass in the inner regions of the Galaxy is baryon dominated and it is possible that baryonic feedback processes might erase the putative dark matter cusp \cite{Bergstrom:2000pn,Bertone:2004pz,Bertone:2010zz}. On top of that, the very  presence of a black hole might disrupt the DM profile, both by making it steeper, through adiabatic contraction \cite{Gondolo:1999ef} or cored through scattering of dark matter particles by stars in the dense stellar cusp observed in the vicinity of the SgrA* \cite{Merritt:2003qk}. Having these uncertainties in mind, we take advantage of the fact that 45 MHz \cite{Guzman:2010da} is a large scale survey and we test its data for a DM contribution farther away from the Galactic Centre, where the DM density is more robustly determined. 

In order to compute the synchrotron signal, the propagation and energy losses of a Galactic electron population need to be modelled.

The propagation of electrons\footnote{DM annihilation produce the same amount of electrons and positrons, so we refer collectively to these particles as 'electrons'. } in the galactic medium is governed by a transport equation, which can be written as \cite{Ginzburg:1990sk}
\bea 
\df{\partial n(x,E)}{\partial t} &=&q(x,E)+\nabla\cdot[K_{xx} \,\nabla n(x,E) - V_c\,n(x,E)]  \nonumber \\
&+& \df{\partial}{\partial p}\,p^2\,K_{pp}\,\df{\partial}{\partial p}\,\frac{1}{p^2}\,n(x,E) \nonumber \\
&-&\df{\partial}{\partial p}\left(\dot{p}\,n(x,E)-\frac{p}{3}\left(\nabla\,V_c\right)\right),
\label{prlong}
\eea
where $n(x,E)$ is the number density of electron per unit energy, and  $q(x,E)$ is the electron source term.  The transport through magnetic turbulence can be described by the diffusion coefficient $K_{xx}$. In the following we will assume, as customary in literature, that the diffusion coefficient is spatially independent and has an energy dependence of the form $K_{xx} =K_0 E^\delta$. $V_c$ is the convection velocity and re-acceleration is described as diffusion in momentum space and is determined by the coefficient $K_{pp}$, and is usually expressed using Alfven speed $v_a$, defined as
\be
K_{pp}\,K_{xx}=\frac{4\,p^2\,v^2_a}{3\,\delta\,(4-\delta ^2)\,(4-\delta)}
\ee

Cosmic rays propagate in the diffusive halo which is usually approximated to have a cylindrical shape with radius $R_h$ around 20 kpc and half-thickness $L_h$ which 
could lie in the range of 1 to 15 kpc. The spatial boundary conditions assume free particle escape, {\em i.e.} {$n(R_h, z, p) = n(r,\pm L_h, p) = 0$}.


{We dedicate the next two sections to describe a semi-analytical and, briefly, fully numerical approach used to derive synchrotron flux. While semi-analytical approach allows in some cases for more physical insight (for instance, we use it to find  the analytical dependence of synchrotron flux on the strength of the magnetic field, see \ref{sec:magfields}), we have checked the robustness of that approach against a full numerical computation, as implemented in the \texttt{GALPROP} code \cite{Porter:2008ve,Vladimirov:2010aq}.}

\subsection{Semi-analytical approach}
For the semi-analytical calculations of the synchrotron signal, following \cite{Baltz:1998xv,Delahaye:2007fr} we neglect re-acceleration and convection terms, which we have checked it is a safe assumption for our particular study\footnote{{However, these two parameters are by default included in the numerical calculation, see the next section.}}, and rewrite Eq.~(\ref{prlong}) as
\bea
\df{\partial n(x,E)}{\partial t} &-& \nabla\cdot[K(x,E) ~\nabla n(x,E)] - \df{\partial}{\partial E}[b(x,E)  n(x,E)] \nonumber \\
&=& q(x,E) ~,
\label{prop}
\eea
where  $b(x,E)$ encodes the energy loss rate. Cosmic ray electrons loose energy mainly through synchrotron radiation and Inverse Compton scattering (IC), with a rate $b(x,E)$ which at the galactic medium is typically of the order $10^{-16}\, \mbox{\rm GeV}\cdot $s$^{-1}$. Additional bremsstrahlung losses of electron energies on the interstellar medium are neglected in this semi-analytical approach, although they may have a very small effect (see for example Fig.\ref{fig:ANvsNUM}), since for the synchrotron frequency we work with, the electron energies of interest are around 1 GeV.

Assuming a steady state, Eq.~(\ref{prop}) can be re-expressed as
\be
\label{prop2}
\df{\partial\tilde n(x,E)}{\partial\tilde t} - K_0~\Delta\tilde n(x,E) = \tilde q(x,E)
\ee   
where the variation with  energy has been parametrized in terms of the parameter  $\tilde t \equiv -\int dE (E^\delta/b(x,E))$. If at the level of propagation one considers that energy losses have an average value over all the diffusion region, i.e. $b(x,E) \approx b(E) = E^2 /\tau$ GeV$\cdot$s$^{-1}$ (see Eq.~(\ref{btot})), then in (\ref{prop2}) $\tilde n(x,E) = E^2 n(x,E)$ and $\tilde q(x,E) = E^{2-\delta} q(x,E)$. The solution of this equation can be found in the Green function formalism to be
\begin{eqnarray}
n(x,E) &=& \df{1}{b(E)}\int_E^\infty dE_s  \\ 
            &\times & \int_{\mathrm{DZ}} d^3 x G(x,E\leftarrow x_s, E_s) q(x_s,E_s) \nonumber
\label{sol}
\end{eqnarray}
where the volume integral is over the diffusion zone (DZ). The Green function $G(x,E\leftarrow x_s, E_s)$ gives the probability for an electron injected at $x_s$ with energy $E_s$ to reach $x$ with energy $E<E_s$, and has a general solution of the form
\bea
G(x,E\leftarrow x_s, E_s) &=& \df{\tau}{E^2} G(x,\tilde t\leftarrow x_s, \tilde t_s) \\
 G(x,\tilde t\leftarrow x_s, \tilde t_s) &=&  \left(\df{1}{4\pi K_0 \Delta\tilde t}\right)^{3/2} e^{-\frac{(\Delta x)^2}{4K_0 \Delta\tilde t}}~,
\label{green}
\eea                   
where $\Delta \tilde t=\tilde t-\tilde t_s$ and $(\Delta x)^2 = (\mathbf x - \mathbf x_s)^2$. The unique argument of the Green function is actually the diffusion length $\lambda = 4K_0\Delta \tilde t$, because the energy dependence enters only in this combination. This is the characteristic length of an electron traveling during its propagation. 
Here we are going to focus on diffusion models for which the half-thickness $L_h\sim 4$ kpc is small compared to the radius of the disk $R_h$, such that in practice the radial boundary has negligible effect on propagation \cite{Delahaye:2007fr}. The
 Green function for which $n(x,E)$ vanishes at $z=\pm L_h$ may be expressed as \cite{Baltz:1998xv} 
\be
\label{greenB}
\tilde G(\lambda, L ) = \df{1}{(\sqrt{\pi}\lambda)^3} \sum_{n=-\infty}^{n=\infty} (-1)^n ~\mathrm{exp}^{-(2 n L +(-1)^n z_s)^2/\lambda^2}~.
\ee 

As for the source term, in the case of production from DM annihilation with cross-section $\langle \sigma v \rangle$, it can be expressed as
\be
q(x,E) = \eta~ \langle\sigma v\rangle \left\{\df{\rho(x)}{m_{DM}}\right\}^2 \df{dN(E)}{dE}~.
\label{source}
\ee
Here $\rho(x)$ is the DM  profile, given in units of GeV/cm$^3$, and $\eta=1/4$ or $\eta=1/2$ depending on the Dirac or Majorana nature of DM. The injection spectrum of electrons is given by $dN/dE$.

With all these ingredients we can express the electron number density as
\be
\label{n}
n(x,E) = \eta~ \langle\sigma v\rangle\left(\df{\rho_\odot}{m_{DM}}\right)^2 \df{1}{b(E)} \int_E^{m_{DM}} dE_s \df{dN}{dE} I(\lambda)~.
\ee
Here $I(\lambda)$ is the so-called  halo function, which is defined has
\be
\label{halo}
I(\lambda) = \int_{\mathrm{DZ}} d^3 x_s \tilde G (x,E\leftarrow x_s,E_s) \left\{\df{\rho(x_s)}{\rho_\odot}\right\}^2.
\ee
When electrons and positrons are created in the Galaxy, they are accelerated by the local magnetic field, thus producing synchrotron radiation, with a flux, per unit frequency $\nu$ per solid angle arriving at position $x$, that is given by
\be
F_\nu =\dfrac{1}{4\pi}\int_{\rm los}dl\int dE_e ~ P(\nu)~ n(x,E_e).
\ee
where integration is performed along the line of sight (los).
Here $P(\nu)$ is the synchrotron energy loss per unit frequency (in GeV/s/Hz), and together with the $n(x,E)$ it must be integrated over all electron energies $E_e$ that produce the same synchrotron at frequency $\nu$. 

In order to express $P(\nu)$ as a function of the electron energy, we use the conservation of energy, $P(\nu) d\nu = (dN_e/dE) P(E) dE$. Here $P(E)$ is an energy-loss rate in units of (GeV/s), and $dN_e/dE$ is the number of electrons with an energy comprised between $E$ and $E+dE$ (not to be confused with $dN/dE$ in Eq.~(\ref{source})). It can be demonstrated that the spectrum of an electron gyrating in a magnetic field has its prominent peak at the energy corresponding to the critical frequency $\nu_c$, $dN_e/dE\simeq \delta(E-E_c)$, where
\be
E_c = 0.5~\sqrt{\dfrac{\nu}{45 ~\mrm{MHz}} \dfrac{10~\mu\mrm{G}}{B}}\,~ \mbox{\rm GeV}.
\label{Ecrit}
\ee 
So 
\[
\int dE_e~ P(\nu)~ n(x,E_e) \simeq \left(\dfrac{dE_e}{d\nu} P(E_e)~ n(x,E_e)\right)_{E_e=E_c}~,
\]
with which finally
\be
F_\nu = \dfrac{1}{4\pi}\int_{\mrm{los}} dl \left(\dfrac{dE}{d\nu} P(E) ~ n(x,E)\right)_{E=E_c}~.
\label{flux}
\ee
Deriving Eq.~(\ref{Ecrit}) with respect to frequency and using Eq.~(\ref{n}) and Eq.~(\ref{flux}), gives the final expression:
\bea
\label{fluxfin}
F_\nu &=& 1.21\times10^8 \dfrac{\mrm{Jy}}{\mrm{sr}} \left\{
\dfrac{\eta}{2}\left(\dfrac{\langle\sigma v\rangle}{3.1\times10^{-26}\mrm{cm}^3/s}\right) \right. \\
&\times&  \left(\dfrac{1\mrm{GeV}/c^2}{m_X}\right)^2 \left(\dfrac{\rho_\odot}{1\mrm{GeV}/c^2/\mrm{cm}^3}\right)^2 \sqrt{\dfrac{1\mu\mrm{G}}{B}} \nonumber\\
&\times&\sqrt{\dfrac{1\mrm{GHz}}{\nu}} \dfrac{P(E_c,B)}{b(E_c,B)} \dfrac{1}{4\pi}\int \dfrac{dl}{\mrm{kpc}}\int_E^{m_{DM}} dE_s \dfrac{dN}{dE}I(E,E_s) ~.\nonumber
\eea
where the synchrotron flux is expressed in Jansky (Jy), $1 Jy = 10^{-26}$ W$\cdot$m$^{-2}\cdot$Hz$^{-1}$ \footnote{In the following, we will express the flux $F_\nu$ in terms of the {\it brightness temperature} of the radiation, $T$, where  $F_\nu = \frac{2h\nu^3}{c^2} [\exp(h\nu/kT)-1]^{-1}$ is the usual black-body relation.}.
In this expression, the synchrotron loss rate is given by
\begin{eqnarray}
\label{bsync}
P(E,B) &=& \df{e^4 E^2 B^2}{6\pi\epsilon_0 m_e^4 c^5}\\
&\simeq& 3.4\cdot 10^{-17}\left(\df{E}{1\mrm{GeV}}\right)^2 
\left(\df{B}{3\mu\mrm{G}}\right)^2\, {\rm GeV}\cdot{\rm s}^{-1}\nonumber
\end{eqnarray}
The total energy loss assumed here is given by 
\be
\label{btot}
b(E,B) = P(E,B) (1 + r_{\mrm{IC}/\mrm{syn}})
\ee 
where 
\be
r_{\mrm{IC}/\mrm{syn}} = \df{2}{3}\df{U_{\mrm{rad}}}{B^2/2\mu_0} \simeq 2 \left(\df{U_{\mrm{rad}}}{8 ~\mrm{eV/cm}^3}\right)
\left(\df{B}{10~\mu\mrm{G}}\right)^{-2}
\label{bICS}
\ee
is the ratio between IC and synchrotron energy loss and
 $U_{\mrm{rad}}$ is the total radiation density.


\subsection{Numerical approach}
In parallel to the semi-analytic expressions, we have determined the synchrotron fluxes using a fully numerical approach as implemented in the \texttt{GALPROP} v54 code \cite{Porter:2008ve,Vladimirov:2010aq}. We ran \texttt{GALPROP} in 2D in galacto-centric cylindrical coordinates ($R$, $z$), solving the CR transport equation on a grid and assuming cylindrical boundary conditions. 
Both re-acceleration and convection are included in \texttt{GALPROP}, though we set $V_c$ to zero in our calculations for simplicity. The energy losses include also bremsstrahlung losses, based on up-to-date whole sky maps of the interstellar gas (for details on generation of gas maps see \cite{FermiLAT:2012aa}). For IC losses \texttt{GALPROP} uses 2D+1 maps\footnote{Two spatial and the frequency dimension.} of the interstellar radiation field, computed based on a model of the radiation emission of stellar populations and further reprocessing in the Galactic dust \cite{Porter:2008ve}. 

The spectrum and distribution of the synchrotron emissivity as seen by an observer at the solar position depends on the form of the magnetic field, and the spectrum and distribution of CR leptons and is computed as a function of $(x,~y,~z,~\nu)$.  The emissivity is then integrated over the line-of-sight to get the synchrotron intensity. We chose the form of the magnetic field consistently with the semi-analytic approach, as detailed below. 


\section{General astrophysical setup}
\label{sec:astro}

\subsection{Astrophysical uncertainties}
\label{sec:astroU} 

Studies coming from N-body simulations have led to popular expressions for the distribution of DM in the Galactic halo, like the Navarro-Frenk-White (NFW) density profile \cite{Navarro:1995iw}
\be
\rho_{\mathrm{ NFW}}(r) = \df{\rho_s}{\df{r}{r_s}\left(1+\df{r}{r_s}\right)^2}~,
\label{NFW}
\ee
or the Einasto profile \cite{Graham:2005xx,Navarro:2008kc}
\be
\rho_{\mathrm{EIN}}(r) = {\rho_s}{\rm exp}\left\{-\frac{2}{\alpha}\left( \left( \frac{r}{r_s}\right)^\alpha-1\right)\right\}~,
\label{einB}
\ee
where $r$ the radial distance from the centre of the DM halo. 
On the other hand, observations of galactic rotation curves as well some of the simulations which include also baryonic feedback on DM density (for a recent example see \cite{Maccio':2011eh}) find DM density profiles which are more cored towards the inner regions of the Galaxy. One example of such profile is the isothermal profile \cite{Begeman:1991iy,Bahcall:1980fb}
\be
\rho_{\mrm{iso}}(r)=\rho_s~ \frac{r_s^2}{r^2+r_s^2}
\label{ISO}
\ee
or modified Einasto profile (in this case the parameter $\alpha$ is smaller when compared to the parameters found in simulations which contain only DM component, \cite{Cirelli:2010xx}).
While the parameter $\alpha$ for the Einasto profile is fixed from a fit to the simulations, the values of parameters $\rho_s$, a typical scale density, and $r_s$, a typical scale radius for the Milky Way are determined from astrophysical observations (e.g. local stellar surface brightness, stellar rotational curves, total Milky Way mass within a given distance..., see e.g. \cite{Bertone:2004pz,Cirelli:2010xx}). 
As we will see in section \ref{sec:DMdensity}, the ratio of the synchrotron signal calculated with DM density of these three profiles gets smaller at higher latitudes, as at those distances (closer to the Solar position) DM density is better constrained.  
{As the rotational curve measurements are poor at distances smaller than 2 kpc (or $\sim 10^\circ$) from the Galacic Center and those regions of the Galaxy are baryon dominated, } we choose the Region Of Interest (ROI) which spans $ |b| \in \left(10\pm 3\right)^\circ$ in Galactic latitudes, and $|l|\lsi3^\circ$ in Galactic longitudes. 

Together with DM density profile, the CR propagation parameters pose one of the main uncertainties in prediction of the synchrotron signal. In \cite{Fornengo:2011iq}, the three models called MIN/MED/MAX \cite{Donato:2003xg} and featured in Table~\ref{CRpar}  are used to probe the uncertainty in CR propagation parameters. Originally, those parameters were derived to produce the maximal, median and minimal anti-proton flux from dark matter, while being compatible with the CR secondary to primary B/C ratio measurement \cite{Maurin:2001sj,Maurin:2002hw}. 
Therefore, by construction, they do not necessarily capture the uncertainty in the electron fluxes in the inner Galaxy, which is of interest here. In section \ref{sec:CRpar} we will comment in more detail on the impact of a choice of a CR parameters, exploring additional sets consistent with the CR data and which were i) derived numerically and ii) shown to reproduce the observed whole sky gamma-ray or radio emission, therefore probing more directly the signals in the inner Galaxy, \cite{FermiLAT:2012aa,Trotta:2010mx,Strong:2011wd}.

{\begin{table}[t]
\begin{tabular}{||c||c||c|c|c||}
\hline\hline
Model        & $L_h$ [kpc]                                                                                & $K_0$ [cm$^2$\,s$^{-1}$]                                          & $\delta$              & $v_a$ [km s$^{-1}$]                  \\
\hline
MIN & 1& $4.8\,10^{26}$ & 0.85 & 0 \\
MED & 4& $3.4\,10^{27}$& 0.70 & 0 \\
MAX & 15& $2.3\,10^{28}$ & 0.46 & 0 \\
\hline
1a & 4 & $6.6\,10^{28}$ & 0.26 & 34.2 \\
1b & 4 & $6.6\,10^{28}$ & 0.35 & 42.7 \\
2a & 10 & $1.2\,10^{29}$ & 0.3 & 39.2 \\
2b & 10 & $1.05\,10^{29}$ & 0.3 & 39.2 \\
\hline
PD & 4 & $3.4\,10^{28}$ & 0.5 & 0 \\
\hline
\hline
\end{tabular}
\caption{Upper three raws: parameter sets derived using a semi-analytical approach, to lead to MIN/MED/MAX anti-proton fluxes at Earth from an exotic Galactic component \cite{Donato:2003xg}. Lower four rows: parameters from \cite{Trotta:2010mx}, consistent with \cite{FermiLAT:2012aa}, derived using \texttt{GALPROP} code in a fit to CR data, and shown to reproduce the gamma-ray diffuse data well. Last row: plain diffusion model, shown to be consistent with the radio data at 22 MHz -- 94 GHz frequencies, \cite{Strong:2011wd}.}\label{tab:CRpar}
\label{CRpar}
\end{table}}

The Galactic magnetic field (GMF) is considered possibly the most important ingredient when dealing with synchrotron radiation. 
In the diffuse interstellar medium it has a large-scale regular component as well as a small-scale random part, both having a strength of order micro-Gauss. The best available constraints in determining the {\it large-scale} GMF are Faraday rotation measures and polarized synchrotron radiation, see e.g. 
\cite{Jansson:2012pc,Pshirkov:2011um,Brown:2007qv,Han:2006ci}, while  {\it random} component is deduced mainly based on the synchrotron emission. 

Several 3D models of the large scale magnetic field are implemented in \texttt{GALPROP} (e.g., \cite{Sun:2007mx,Sun:2010sm}). However, as our ROI is away from the Galactic disk and the  small-scale random component is expected to dominate the total value of the magnetic field, we use a simple parametrisation 
for a \emph{total} magnetic field as customary in literature:
\be
B(\rho,z) \propto \mrm{exp} \left( - \df{r -r_\odot}{R_m} - \df{|z|}{L_m}\right)~.
\label{Bexp}
\ee
The parameters $R_m$ and $L_m$ should in principle depend on the diffusion model assumed (or {\em vice versa}), since the propagation in the Galactic medium is intimately related with the magnetic field. We follow \cite{Fornengo:2011iq}  in taking $L_m =\delta\cdot L_h$ and $R_m = \delta\cdot R_{h}$. It was shown in that work that an actual extent of $L_m$ and $R_m$ does not play a critical role (the difference between a constant magnetic field and the exponential form defined above is $\lsi 30\%$), as long as the field extends into our ROI (i.e. $L_m\gsi 1$ kpc). The normalisation at Sun's position $B_\odot$ is more or less well constrained, and we take the value of $6~\mu$G \cite{Strong:2004de}, consistent with the measurements \cite{Jansson:2012pc,Pshirkov:2011um,Brown:2007qv,Han:2006ci}. However the value of the field in the inner Galaxy is considerably less known and we rewrite the normalisation of (\ref{Bexp}) as \footnote{following \cite{Dobler:2011mk}.}
\be
\label{Bprofil}
B_0 \equiv B_\odot [1 + K~ \Theta(R_{IG} - r)] 
\ee
where $\Theta(R_{IG}-r)$ is the unit step function as a function of $r=\sqrt{\rho^2 + z^2}$. With this change, while leaving $B_\odot$ unchanged \emph{locally} we allow for the magnetic field to have a higher effective normalization $B_\odot~(1+K)$ in the inner Galaxy. We arbitrary set $R_{IG}=2$ kpc, in order to cover our ROI. 
We therefore fix the spatial dependence of the B field and explore the impact of overall normalisation $B_\odot~(1+K)$ in the inner Galaxy, on synchrotron fluxes in section \ref{sec:magfields}.

Contributions to the energy losses for electrons are assumed to come only from synchrotron and IC processes in our semi-analytical approach, as commented above. In principle, the radiation density $U_\mrm{rad}$ (see Eq.~(\ref{bICS})) has a spatial profile, which affects the synchrotron flux estimations. However, in the semi-analytical estimations used here we take $U_{\mrm{rad}}$ to be constant, which turns out to be a good approximation for the particular study performed in this work, as is justified in the next section. For the value of $U_{\mrm{rad}}$ we take 8 eV cm$^{-3}$, which is approximately the value read off the ISRF maps used by \texttt{GALPROP} in our ROI. 
\footnote{
Note however that the conditions in the inner Galaxy might be quite different from a simple CR propagation setup assumed here. In particular, observations of the bubbles like structures Centerd at the Galactic centre and extending to 50 deg in latitudes in gamma rays \cite{Su:2010qj,Su:2012gu} and WMAP haze at microwave frequencies, observed by WMAP \cite{Finkbeiner:2003im} and Planck \cite{planckhaze}, witness of possibly more complicated configuration of the magnetic fields and CR propagation parameters in that region. While bubble-like structures appear sub-dominant with respect to the standard components of the diffuse emission in our ROI and electron energies of interest,  we caution that before their origin is understood, the actual structure of magnetic fields or the CR propagation conditions, cannot be reliably modeled.}

Concerning the frequency of observation we focus on the data taken at 45 MHz. In \cite{Fornengo:2011iq} 
it was shown that in a wide range of frequencies (22 to 1420 MHz) the change in the measured synchrotron flux is very small, while going to lower frequencies maximises a synchrotron signal of low mass WIMPs. The RMS temperature noise of this survey is 3500 K, which is generally subdominant with respect to the systematics errors involved in theoretical modeling. 


\subsection{Cross-check of semi-analytical calculation of synchrotron fluxes}
\label{sec:cross} 

\begin{figure}[ht]
\includegraphics[width=0.5\textwidth,angle=0]{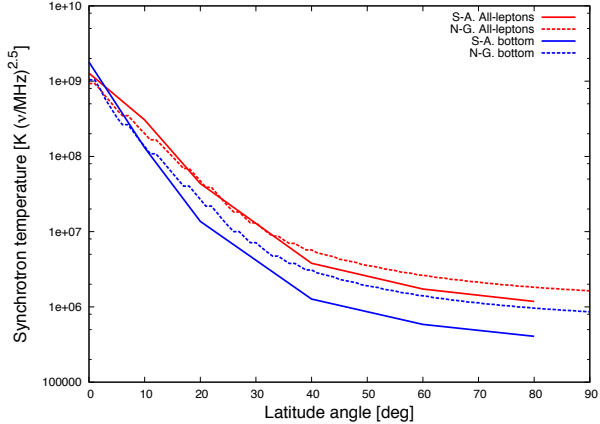}
\caption{\footnotesize{Comparison between semi-analytical (S-A.) and numerical/Galprop (N-G) approaches, for different annihilation channels{, and DM mass of $m_{DM}=10$ GeV}. A frequency of 45 MHz was used, assuming a magnetic field normalised to $B_{\mathrm{GC}}=10~\mu$G {(which corresponds to $K$=0, i.e. no enhancement of the magnetic field in our ROI)}. }}
\label{fig:ANvsNUM}
\end{figure}

In Fig. \ref{fig:ANvsNUM} we show a comparison of synchrotron spectra obtained with the semi-analytical  (SA) and numerical/\texttt{GALPROP} (NG) approaches. We see in general a good agreement between the two predictions. We observe that in the region of low latitudes, SA calculation gives relatively larger fluxes than the NG counterpart, while for the region of large latitudes, it is the opposite. That is expected, as in the SA approach we have used a constant value for the radiation density
, while in the NG case, a full space dependent profile for $U_{\mrm{rad}}$ is used. Typically this radiation density gets smaller as the distance from the galactic centre increases, so the synchrotron flux becomes less suppressed, explaining why the \texttt{GALPROP} estimations are typically larger for large angles and vice versa, see Fig. \ref{fig:ANvsNUM}.  

However, as commented in section D, the data of synchrotron flux we used to constrain dark matter models is the one taken at a latitude of $\sim 10^\circ$, and for this latitude, the semi-analytical and numerical approaches are in very good agreement. In other words, for this latitude the approximation of taking $U_{\mrm{rad}}$ constant works very well.

Finally, we have checked a good agreement with the results of  \cite{Fornengo:2011iq} using their particular setup. The biggest difference between our works is that we assume a factor of $\sim 2$ higher ISRF energy densities, inspired by the values implemented in the \texttt{GALPROP code} in our ROI. That in turn implies higher overall energy losses, and therefore lower synchrotron fluxes (by about the same factor, {if IC losses are the dominant ones, see Eq.~(\ref{FofB}})), for the otherwise same parameter set.

\subsection{Synchrotron signal for different choices of DM density profile} \label{sec:DMdensity}

\label{sec:DMprofiles}

In the remainder of the text we will focus on two DM profiles, NFW, eq.~(\ref{NFW}) and isothermal (ISO) Eq.~(\ref{ISO}), using the following values of parameters, consistent with observations: $\rho_s = 0.31$ GeV\,cm$^{-3}$, $r_s=21$ kpc, for an NFW profile, and $\rho_s=1.53$ GeV\,cm$^{-3}$ and $r_s=5$ kpc for Isothermal profile\footnote{Note that we make a conservative choice by choosing a rather extended core. Smaller values of $r_s$ would result in fluxes more similar to those obtained with the NFW profile.}, \cite{Fornengo:2011iq}.  However, in this section we also show the prediction for the synchrotron signal in the case of a modified Einasto profile. In particular, this profile is modified to have a shallower inner slope than the usual Einasto profile found in DM-only simulations and it describes better results of simulations which include baryonic feedback (parameters we use are $\alpha =0.11$, $r_s=35.24$ kpc, $\rho_s=0.041$ GeV cm$^{-3}$ \cite{Cirelli:2010xx}). As this profile is 'bulkier' at distances ~1 kpc from the GC, the DM signals are generally higher than those of NFW in that region.

\begin{figure}[ht]
\includegraphics[width=0.5\textwidth,angle=0]{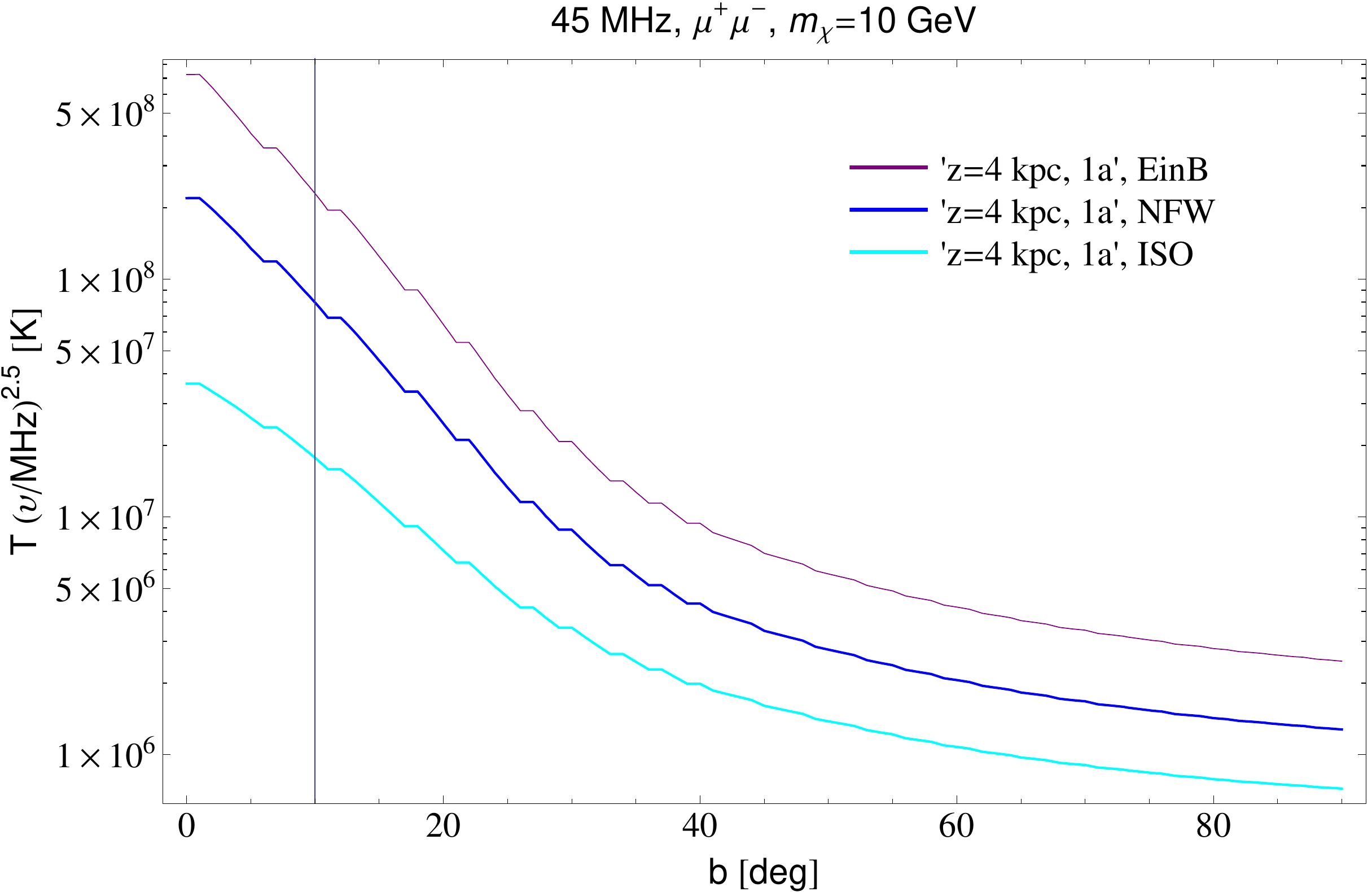}
\includegraphics[width=0.5\textwidth,angle=0]{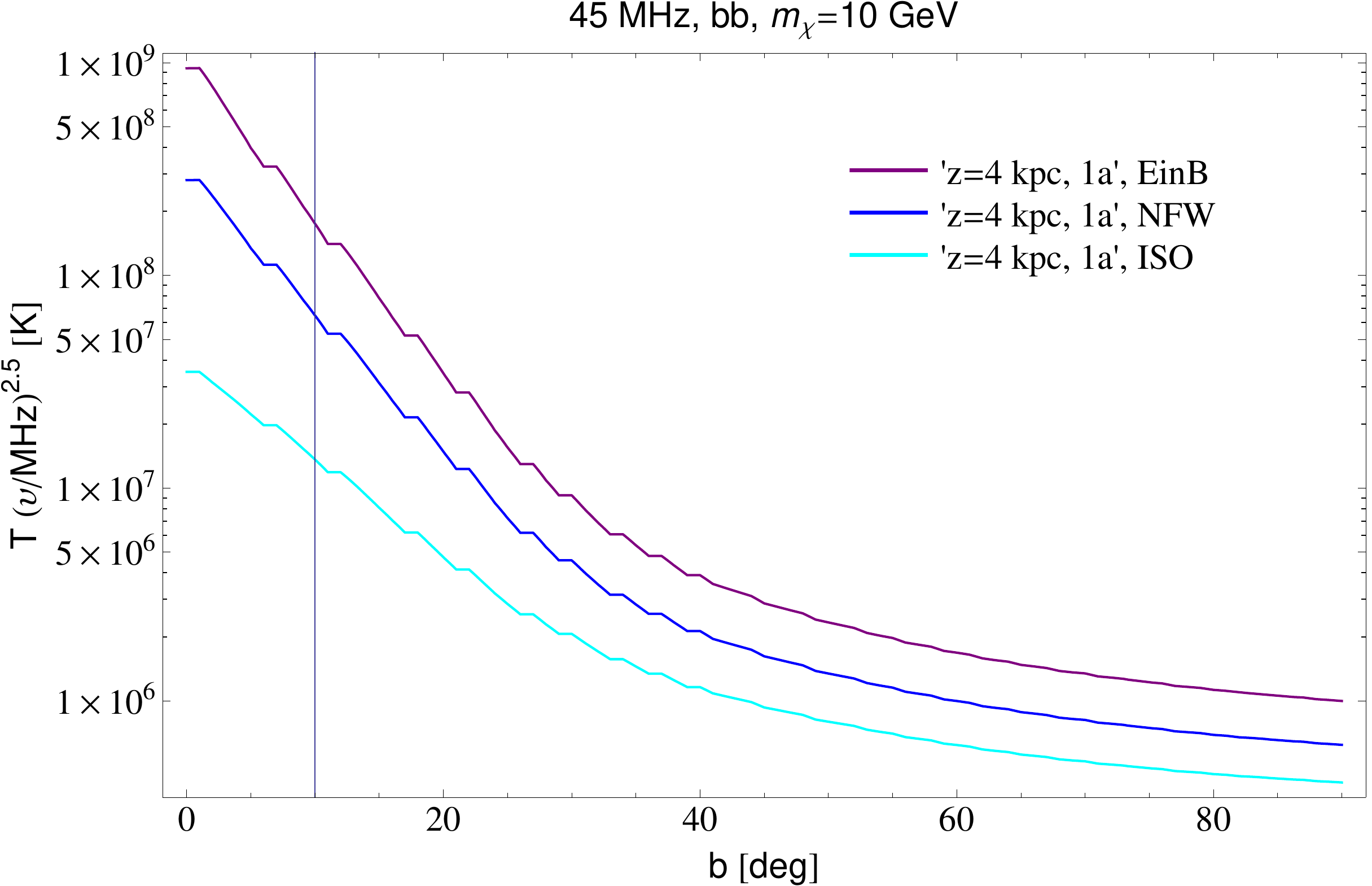}
\caption{\footnotesize{Comparison between synchrotron signals for a DM mass of 10 GeV annihilating to muons (top) and b quarks (bottom figure) for three DM density profiles: modified Einasto, NFW and ISO thermal profile. $\rho_{\odot} = 0.43$ GeV\,cm$^{-3}$ is assumed for this plot (see text for the remaining parameters) and propagation of electrons is done using a CR propagation setup as shown in Table \ref{tab:CRpar}.}}
\label{fig:DMprofiles}
\end{figure}

In the parameter sets we chose, the local value of DM density is set to $\rho_{\odot} = 0.43$ GeV\,cm$^{-3}$. One should also keep in mind that the overall normalisation of DM distribution $\rho_{\odot}$  is uncertain, being in the $(0.43 \pm 0.113 \pm 0.096$) GeV\,cm$^{-3}$ range\footnote{However, depending on the analysis, the uncertainty window on this value can vary in the 0.2 -- 0.8 GeV cm$^{-3}$ range.} \cite{Salucci:2010qr}. Therefore, in addition to the differences in the signal caused by the DM profile shape, and shown in Figure \ref{fig:DMprofiles}, synchrotron signals scale with $\rho^2 _{\odot}$.


\subsection{Synchrotron signal for different choices of CR parameters} \label{sec:CRpar}
\label{sec:CRpars}

As discussed above, MED/MIN/MAX sets of CR propagation parameters were derived using a semi-analytical description of CR propagation, and a fit to B/C measurement, with a requirement to produce minimal, medium, and maximal DM generated anti-proton fluxes, at a Solar position. These models were recently reanalysed in \cite{Bringmann:2011py}, where  whole sky radio data together with B/C measurements where considered, testing therefore impact of these parameters on electron population, and its synchrotron emission in various regions in the Galaxy. This work concludes that both MIN and MAX models are disfavored by radio data towards the galactic anti-centre. Demanding consistency with B/C nuclei data, it is found that small halo sizes $L_h\sim$ 1 kpc are essentially excluded (see also \cite{Strong:2011wd}), but also large values $L_h \gsi$ 15 kpc show some tension with radio data, arguing that MIN/MAX sets of parameters present somewhat extreme choices when compared to observations complementary to CR nuclei data.

\begin{figure}[ht]
\includegraphics[width=0.5\textwidth,angle=0]{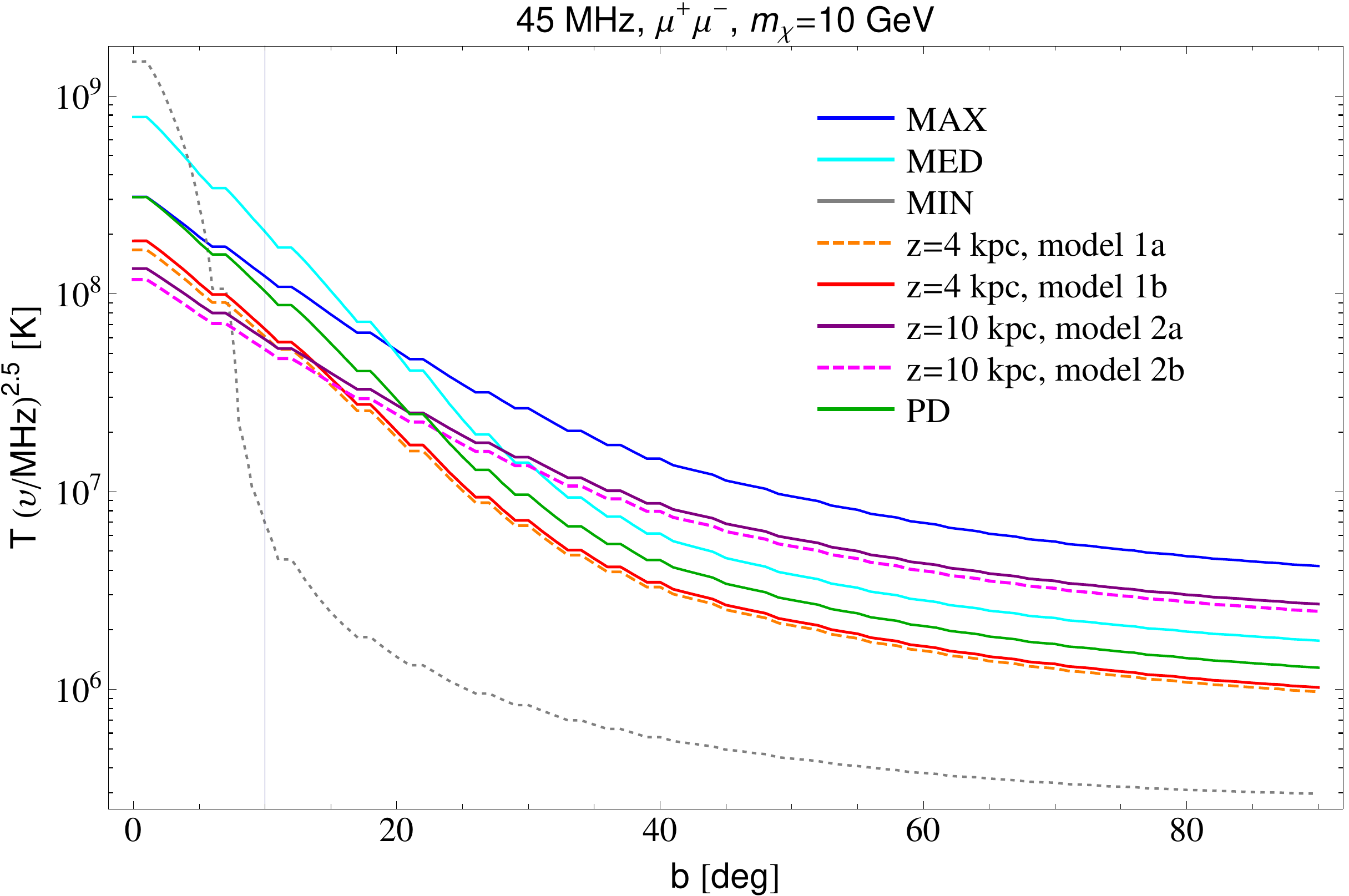}
\includegraphics[width=0.5\textwidth,angle=0]{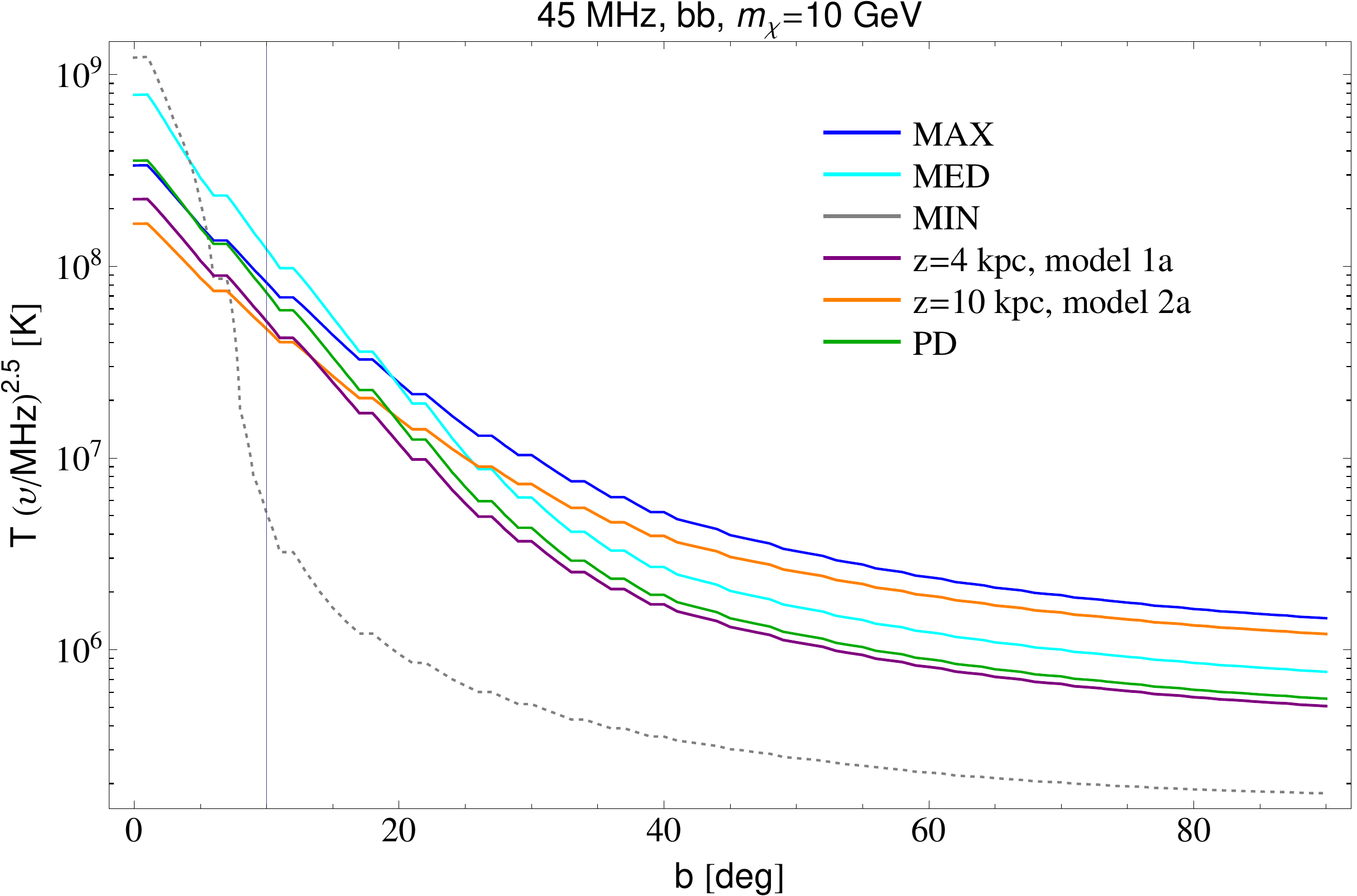}
\caption{\footnotesize{Comparison between synchrotron signals for a DM mass of 10 GeV annihilating to muons (top) and b quarks (bottom figure) for different CR propagation setups, detailed in Table \ref{tab:CRpar}. NFW DM profile is assumed here.}}
\label{fig:CRdiffmodels}
\end{figure}

In parallel to the above mentioned analysis based on semi-analytical approaches, we also use results of three analysis based on a numerical calculation with the \texttt{GALPROP} code \cite{FermiLAT:2012aa,Trotta:2010mx,Strong:2011wd}. 
In \cite{FermiLAT:2012aa} the \texttt{GALPROP} code is used to derive a set of CR parameters which provide a good fit to the CR data, and at the same time reproduce the gamma-ray data well. In a parallel work, \cite{Trotta:2010mx},  a full Bayesian analysis of a fit of CR models is made using \texttt{GALPROP} to confront CR data and to derive the best fit model and its scatter. It is shown that gamma-ray predictions of such model are consistent with the {\it Fermi} LAT gamma-ray measurement. We will use the best fit model from \cite{Trotta:2010mx}, and few models within a 2 sigma scatter of the Bayesian analysis, see Table \ref{tab:CRpar}. We show a comparison of synchrotron signal calculated with these choices of CR propagation parameters in Fig.\ref{fig:CRdiffmodels}. The \texttt{GALPROP} based parameter sets in general have lower values of parameters $\delta$ and consequently higher values $K_0$ are obtained in a fit to B/C data. That in turn means that diffusion is higher in this set of models, {\em i.e.} electrons diffuse out of our ROI, and contribute instead more at higher latitudes, as seen in Fig.  \ref{fig:CRdiffmodels}. 

A study in \cite{Strong:2011wd} has shown, however that some models with re-acceleration ($v_a\ne 0$, similar to the case 1a above) are in tension with the large set of radio data. The main reason for this is that production of CR secondary electrons is enhanced in this case and produces too strong radio emission at low frequencies. We therefore consider also a plain diffusion model (PD) often used in literature and shown to be consistent with the radio data in \cite{Strong:2011wd}.

In the remainder of the text we will adopt the MED model together with the NFW profile and $\rho_\odot=0.43$ GeV cm$^{-3}$ as an moderately optimistic scenario, and the '1a', \texttt{GALPROP} based CR propagation model, together with the Isothermal profile and  $\rho_\odot=0.3$ GeV cm$^{-3}$ as a moderately conservative setup. 

\subsection{Synchrotron signal for different magnetic field choices} \label{sec:magfields}
\label{sec:magn}

In this section we study how DM limits change depending on the assumptions of the overall normalisation of the magnetic field in our ROI. It has already been noticed (see e.g. \cite{Boehm:2010kg}) that for a fixed electron injection spectrum there exists an optimum value for the magnetic field which maximises the synchrotron flux at a given synchrotron frequency. However in this section we want to understand this fact in more detail. 

As can be seen in Eq.~\ref{fluxfin}, the magnetic field influences the flux through the energy losses, and through the electron energy.  In principle, the electron spectrum $dN(E_s)/dE$ affects the maximisation of the flux with respect to the magnetic field $B$. A way to see this is the following: in the assumption of considering a one-to-one correspondence between the emitted synchrotron frequency $\nu$ and the electron energy $E_c$, (see Eq.~\ref{Ecrit}), 
for a given value of the magnetic field $B$, the correspondent $E_c$ could be disfavored (or not available) by the electron spectrum produced by the DM. 
However, as in this study we are going to work with synchrotron data with frequency of $45$ MHz, and magnetic fields $\gtrsim 6~\mu$G, 
the electron energies producing those desired frequencies are always smaller than 1 GeV, which can perfectly be produced by our DM candidates in the range [1-200] GeV. So in practice, for our analysis, the only dependence on $B$ that matters are encoded in the energy losses. Indeed, given a specific frequency $\nu$, and a given annihilation channel, it can be expressed in the following form:

\be
F(\nu,B) \propto \left(\df{B^2}{\alpha + B^2}\right) \df{1}{\sqrt{B}}~,
\label{FofB}
\ee 


where $\alpha$ represents here the rest of energy-losses, here assumed to be only IC. Note that since both synchrotron and IC losses scale with energy as $E^2$, the energy dependence cancels in this particular analysis. From (\ref{FofB}) one observes two extremal cases: one in which synchrotron loss is negligible with respect to the rest of energy losses, for which the flux scales with $B$ as $F\sim B^{3/2}$, thus increasing as $B$ increases, and the other, in which synchrotron is actually the dominant energy loss, after which the flux scales as $F\sim 1/\sqrt{B}$, thus decreasing as $B$ increases. In other words, there will be an intermediate value of the magnetic field for which synchrotron becomes the dominant energy loss, and this value is actually the one maximising the flux. 

\begin{figure}[ht]
\hspace{-1.5cm}
\includegraphics[width=0.55\textwidth]{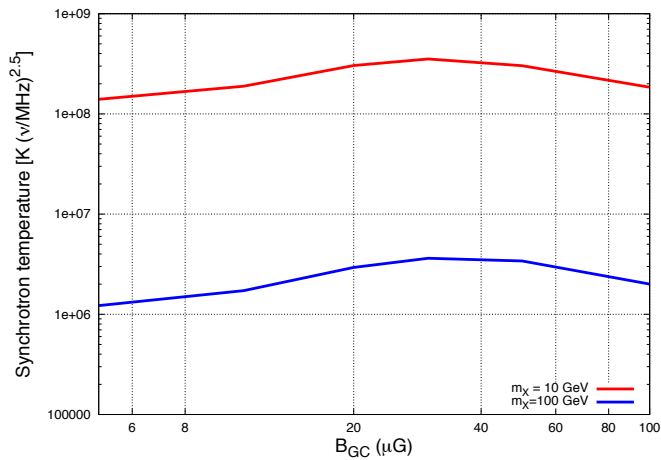}
\caption{\footnotesize{Flux predicted at 10 deg off the GC, as a function of the value of magnetic field at the GC (roughly, at our ROI, we have $B_{\mathrm{ROI}}\approx 0.5 B_{\mathrm{GC}}$). Lines represent the results in the Semi-Analytical approach. We assume a DM annihilating directly to electrons, using the NFW profile and the MED diffusion model}}
\label{fig:magfield}
\end{figure}

Figure \ref{fig:magfield} shows the shape of synchrotron flux as a function of the value of magnetic field at GC. Taking into account only IC (apart from synchrotron of course), one can have an idea about the maximum of the flux already by direct differentiation of (\ref{FofB}), assuming the values of $\alpha$ correspondent to this case. The value of $B$ for which the flux is maximal scales as $B_{\mathrm{GC}}^{\mrm{max}}\propto\sqrt{U_{\mrm{rad}}}$. For $U_{\mrm{rad}}=8$ eV/cm$^3$, $B_{\mathrm{GC}}^{\mrm{max}}\simeq 26~\mu$G.




In the following we will show DM limits using two cases for magnetic field configurations. The first one, a standard $B_{\mathrm{GC}} \simeq 10~\mu$G value in our region of interest is assumed, and the second case will be the normalisation which maximises the synchrotron signal, {\em i.e.} $B_{\mathrm{GC}}\simeq 26~\mu$G.

\section{Other indirect constraints}
\label{section:dwarfs}

In the following we will compare the limits we derive from considering radio data to i) those derived by the {\it Fermi} LAT collaboration from the non-observation of dwarf spheroidal Galaxies (dSphs) in gamma-rays \cite{Ackermann:2011wa} and ii) constraints  based on the measurement of the CMB anisotropy spectra \cite{Galli:2011rz}. These limits are {among}  the strongest to date and uncertainties involved in their calculation are complementary to the ones derived in this work.

The main strength of considering dSphs is that they are DM-dominated and lack active astrophysical production of gamma-rays. {We will refer to an} analysis of the {\it Fermi} LAT data {which}, for the first time, combines multiple Milky Way satellite galaxies in a single joint likelihood fit and includes the effects of uncertainties in $J$ factors (line of sight integrals of DM density squared, which define the DM annihilation rates). The improvements of the limits derived in {such an} analysis over 10 years of {\it Fermi} LAT mission was estimated in \cite{Cotta:2011pm}. In the low-energy regime, the sensitivity increases as roughly the square-root of the integration time. However, in the high-energy (limited background regime) the LAT sensitivity increases more linearly with integration time. Thus, 10 years of data could provide a factor of $\sqrt{5}$ to 5 increase in sensitivity. Additionally, new optical surveys (such as Pan-STARRS\footnote{http://pan-starrs.ifa.hawaii.edu/public/} and the Dark Energy Survey\footnote{http://www.darkenergysurvey.org/}) could provide a factor of 3 increase in the number of detected dSphs corresponding to an overall increased constraining power $\sqrt{15}$ to 15. As we focus mainly on DM masses $\lsi 100$ GeV, we will use $\sqrt{15}$ to illustrate potential improvement in our plots. 

In order to translate these limits in terms of democratic couplings to leptons and hadrons (shown in Figures~ \ref{fig:svVSmX1} and \ref{fig:svVSmX2}) we use i) the limits to $\tau$ channel worsened by a $1/3$ branching to this channel; dSphs constraints on $\mu$ are significantly weaker (this likely holds true also for the $e$ channel, however these limits were not published) and ii) limits on $b$ quark states, as the gamma-ray spectra are very similar for all quark channels.

CMB constraints arise from redshifts in the range $100 \lsi z \lsi 1000$ \cite{Padmanabhan:2005es,Slatyer:2009yq,Galli:2009zc,Hutsi:2011vx,Galli:2011rz}. The physical effect of energy injection of (exotic) particles around the recombination epoch is that it results in an increased amount of free electrons, which survive to lower redshifts and increases the width of last scattering surface, consequently suppressing  the amplitude of some of the oscillation peaks in the temperature and polarisation CMB power spectra. As constraints come from high redshifts well before the formation of any sizeable gravitationally bound structure, this set of constraints does not depend on highly uncertain parameters related to structure formation. Detailed constraints have been recently derived in  \cite{Galli:2011rz}, based on the WMAP (7-year) \cite{Komatsu:2010fb} and Atacama Cosmology Telescope 2008 data \cite{Fowler:2010cy}. The constraints are somewhat sensitive to the dominant DM annihilation channel: annihilation modes for which a portion of the energy is carried away by neutrinos or stored in protons have a lesser impact on the CMB; on the contrary the annihilation mode which produces directly $e^+e^-$ is the most effective one. The limits on hadronic channels were not derived in this work. We therefore show CMB limits only for the democratic coupling to leptonic states, in which case we combine the published limits to $\mu$ and $e$ channels, renormalised by  a branching factor of $1/3$.   We also show the projected sensitivity of the PLANCK telescope in the near future, as calculated in \cite{Galli:2011rz}.

\section{Constraints on Dark Matter Models}
\label{sec:models}
\subsection{Effective operators}
\label{sec:effectiveOperators}

\subsubsection{Mono-events at colliders}
\label{section:collider}

The very same interactions responsible for DM annihilation in the galactic medium that ultimately can produce synchrotron fluxes, are the ones by which a DM signal could be produced and measured in collider experiments. 
At LEP, for example, when a pair of electron-positron collides at high energy  a plethora of final states are produced.
 Among these states, the production of a DM candidate $\chi$ can be characterised by missing energy and the emission of a single photon
 \cite{Abdallah:2003np,Abdallah:2008aa}. {These data can then be used to constrain theoretical models. We first choose to use the powerful machinery of effective field theory (EFT) to capture features from a broad class of WIMP models. In particular we assume that at the relevant energies, the only available degrees of freedom are the dark matter particle itself and the Standard Model. Provided this is true for all energies of interest, the EFT provides a common language which allows one to compare the constraints from different types of experiments \cite{Beltran:2010ww}.
 } Introducing new physics scales $\Lambda_i$ one can define effective couplings between ({\em e.g.} fermionic) DM  to  SM fermions $f$ 
 by
 
 \be
\mathcal O^f_i = \df{1}{\Lambda_f^2} (\bar\chi\Gamma_i\chi)(\bar f \Gamma^i f),
\hspace{0.5cm} \mathcal O^f_t = \df{1}{\Lambda_f^2}(\bar\chi f)(\bar f\chi)~,
\label{effop}
\ee
 
\noindent
where $\Gamma_i$ =($1;\gamma_\mu$; $\gamma_5\gamma_\mu$) for scalar ($\mathcal O^f_S$), 
vector ($\mathcal O^f_V$) or axial ($\mathcal O^f_A$) operators respectively. 
Performing a Monte Carlo simulation studying the signal ($e^+ e^- \rightarrow \gamma \bar\chi\chi$) plus
 background (e.g. $e^+ e^-\rightarrow \gamma\bar\nu\nu$) processes, one can compare the theoretical background predictions with the real data produced
  at LEP \cite{Fox:2011fx} and put lower bounds on the effective coupling $\Lambda_e$. 
 This effective approach can be understood as a limit of microscopic models with heavy mediator (Higgs boson or extra $U(1)$ $Z'$
 for instance), discussed in sections \ref{sec:HiggsPortal} and \ref{sec:ZpPortal}.

A similar study has been made \cite{Bai:2010hh,Goodman:2010ku} using results from the CDF collaboration with Tevatron \cite{CDF}, and more recently \cite{Fox:2011pm} 
from ATLAS study \cite{Aad:2011xw}, and by the CMS Collaboration itself \cite{Chatrchyan:2012te}. The analogous types of operators defined in Eq.~\ref{effop} 
are used as source of processes like ($\bar q q\rightarrow j \bar\chi\chi$), where $q$ are SM quarks and $j$ stands for a single jet. A mono-jet final
 state is then used to put constraints on $\Lambda_q$ and thus on (in this case) hadronic annihilation cross-section, in the very same way as in \cite{Fox:2011fx}.

\begin{figure}[ht]
\includegraphics[width=0.5\textwidth]{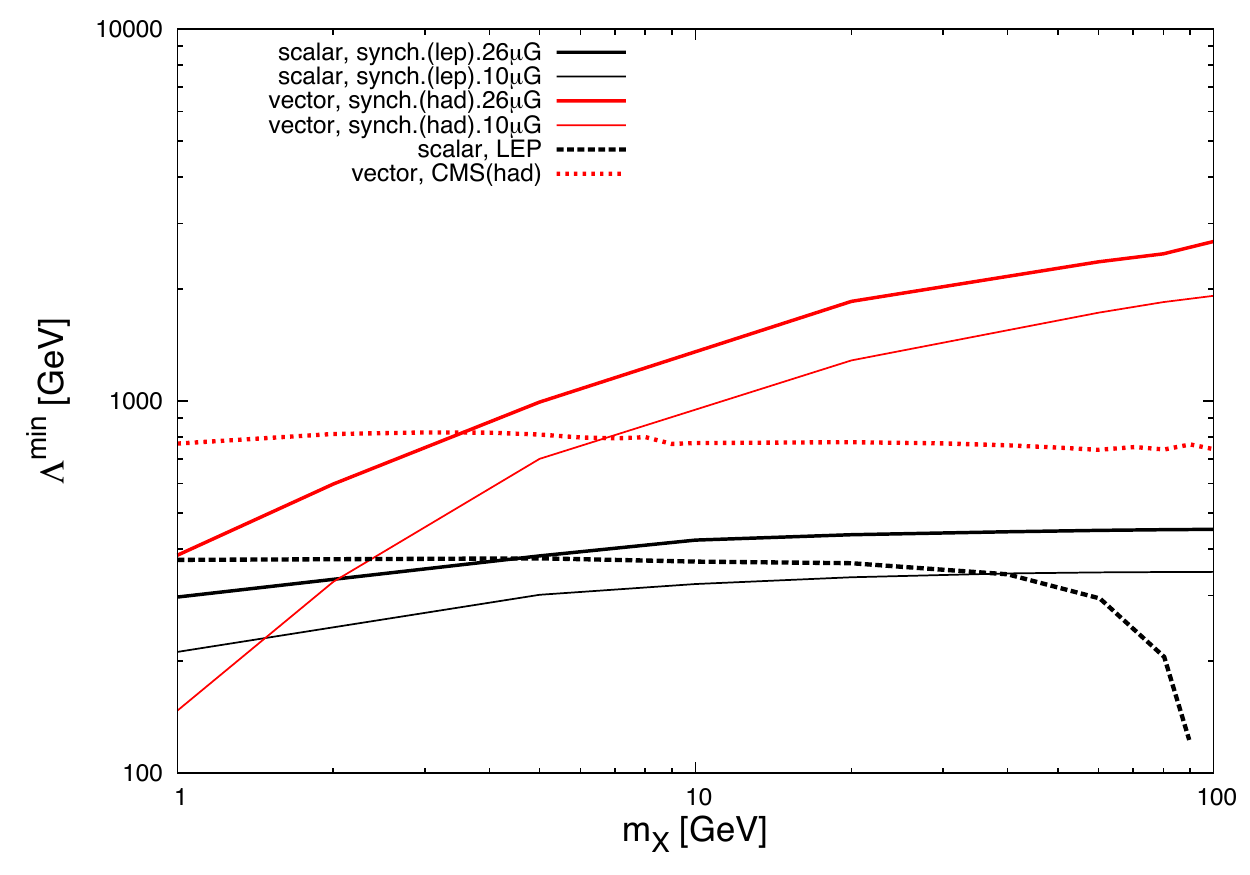}
\caption{\footnotesize{ Lower bounds on the effective scales $\Lambda_{l,q}$ from some collider studies, as well as from synchrotron data for a certain choice of parameters (see text for details). ``CMS(had)" means CMS bounds to quark operators. Solid lines are bounds from synchrotron radiation data, while dashed lines are bounds from LHC (CMS) or LEP. Red lines correspond to quark operators while black lines correspond to leptonic ones.}}
\label{fig:LEPLHC}
\end{figure} 

 As an illustration, we show in Fig.~\ref{fig:LEPLHC} the limits we obtained on the leptonic and hadronic effective scales,
 $\Lambda_l$ and $\Lambda_q$,
  obtained by LEP and CMS as function of the DM mass $m_{\chi}$ in the case of scalar/vectorial interaction
  with universal leptonic/hadronic coupling.
 We notice that the bounds on $\Lambda_{f=l,q}$ become weaker as DM mass increases because of the phase space reduction,
  up to some point in which it is kinematically impossible -given the energies of the experiment- to produce the pair of DM particles
  (the threshold is of course higher in LHC experiments than at LEP). 
The lower bounds on $\Lambda_l(q)$ is around 300 GeV (600 GeV) and relatively independent of the DM mass as one
measures missing energy. It becomes then interesting to compare this limit to the one
derived from synchrotron emission.
 These bounds obtained by the LEP and CMS can also be converted directly into upper bounds on the leptonic/hadronic annihilation 
 cross-section $\langle\sigma v\rangle_{ee/qq}$  from
  the inverse processes ($\bar\chi\chi\rightarrow \bar e e/qq$), generated by the operators defined in Eq.~\ref{effop} (see Appendix \ref{AppA} for details).

\subsubsection{Synchrotron {\em vs} Collider {and complementary Indirect Detection} bounds}

In order to model the synchrotron signal of a standard astrophysical origin, a HASLAM 408 MHz map \cite{Haslam:1982zz} can be used as a template. It is a full sky radio map with the best angular resolution and sensitivity. 
Indeed, a 408 MHz frequency is well suited to gauge the contribution from the harder population of galactic electrons, 
while leaving lower frequency maps sensitive to test a possible contribution from softer electrons originating in annihilations 
of light dark matter candidates. 

\noindent
In \cite{Fornengo:2011iq} it has been shown that by extrapolating  the HASLAM data down to 45 MHz, to model astrophysical 
synchrotron emission, one is left with $\sim30\%$ residuals when compared to the actual 45 MHz data. Those residuals do not have 
the proper morphology of a DM signal and therefore no DM detection could be claimed. Based on this analysis it seems reasonable to assume that the 
current uncertainty ({\em i.e.} systematics) in modelling the astrophysical emission is at a level of  $\sim30\%$. 
However, in our analysis, we do not attempt to model specific astrophysical signals. Instead, we decide
to apply a conservative 95\% CL DM limits without assuming any contribution from astrophysical background,
  and which we label $no$ $bckgd$, in the figures
  and also show how this limits change as a function of the systematic uncertainties on the astrophysical model,
  expressed as a background uncertainty in \% of the data.
  
  Our procedure is to compute the synchrotron flux coming from DM ($F_{\mathrm{DM}}$) in our ROI, and then constrain the result with the available data corresponding to that region, $F_{\mathrm{obs}}$, by requiring that $F_{\mathrm{DM}} \leq F_{\mathrm{obs}}+2\sigma$, where $1\sigma$ is the uncertainty considered in each case (see Figures \ref{fig:backgr1} and \ref{fig:backgr2}). In the case of $no~bckgd$, $1\sigma$ is the rms temperature noise, taken to be 3500 K.  
  
  For illustration, in Figure~\ref{fig:LEPLHC} we have included the bounds on $\Lambda_i$ obtained from synchrotron data
  at 45 MHz with a 26  and 10 $\mu G$ magnetic field in our ROI, with the hypothesis that the dark matter signal lies within 5\% of uncertainty
  of the background, and using the NFW+MED set-up. We clearly see that the synchrotron radiation, with the present data, can already give stronger limit
  on the effective scales, independently on the value of the magnetic field, and is complementary  to the bounds from accelerator 
  searches. Indeed, whereas there are no "threshold effect"
  for synchrotron radiation at large DM mass, one notice that for low masses, in the hadronic channel the bounds becomes weaker. We comment on this behaviour below. 



\begin{figure}[ht]
\includegraphics[width=0.5\textwidth,angle=0]{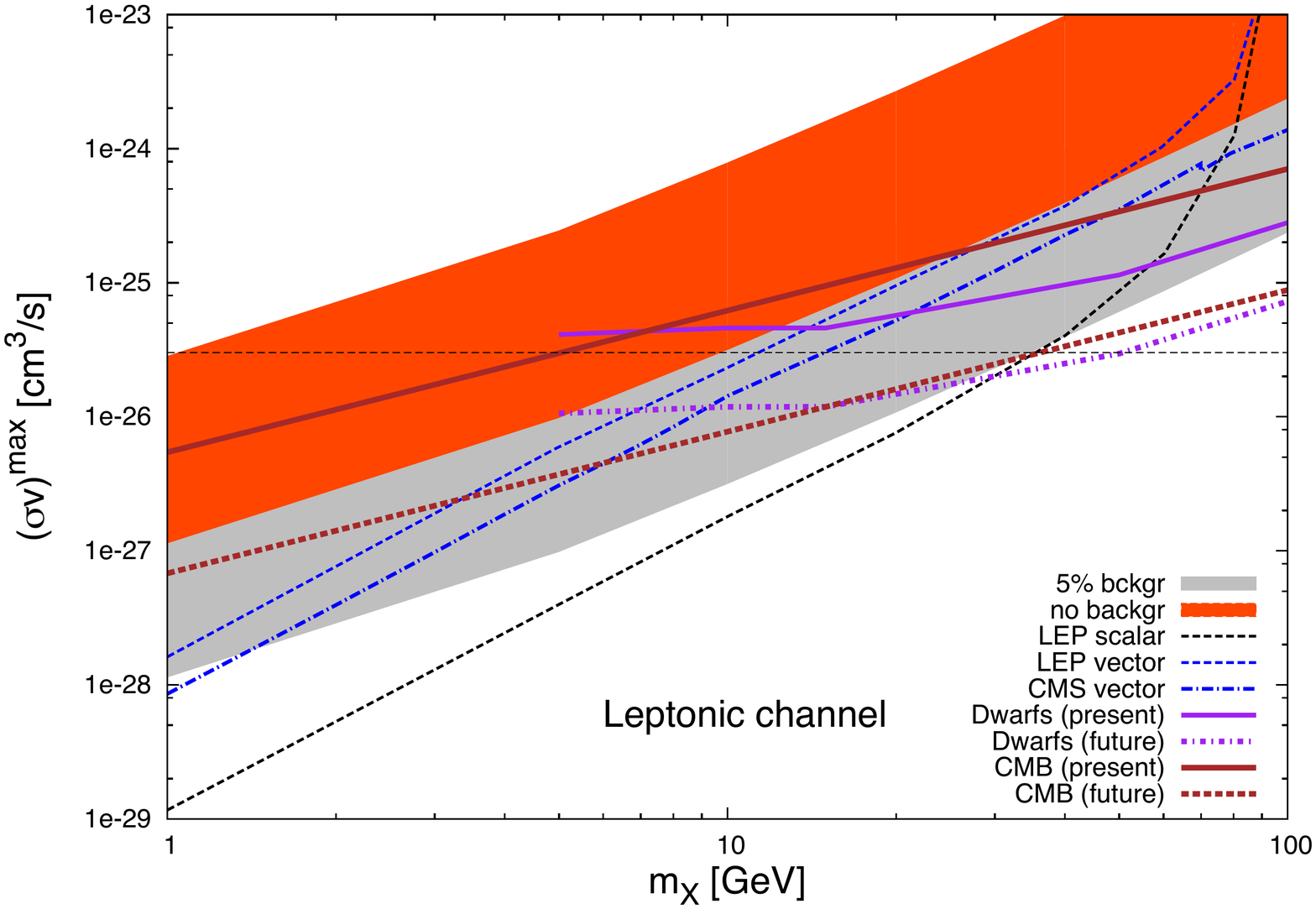}
\includegraphics[width=0.5\textwidth,angle=0]{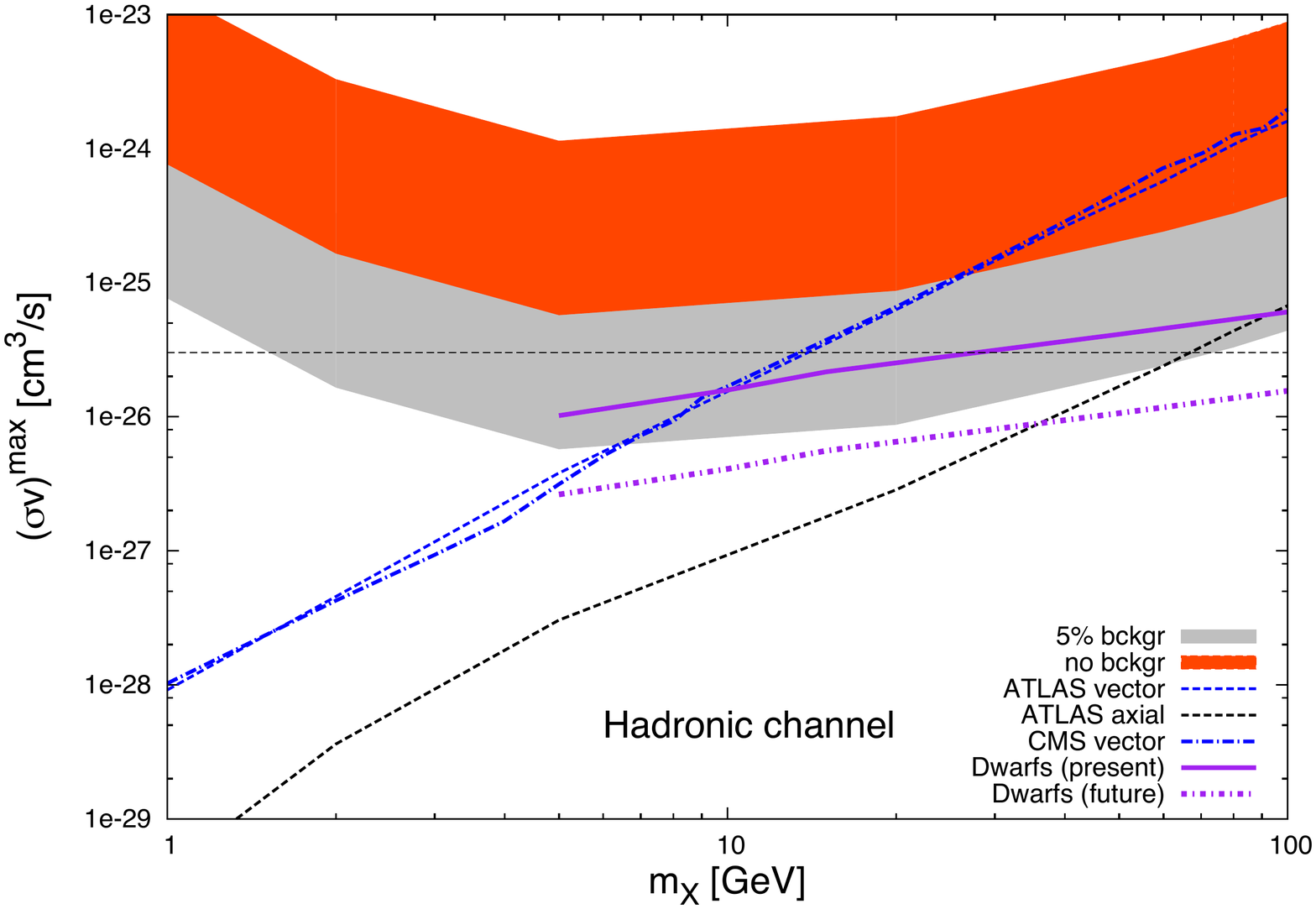}
\caption{\footnotesize{ Synchrotron bounds on $\langle\sigma v\rangle$, assuming DM couples democratically to all charged-leptons (left panel) or to all kinematically available quarks (right panel). In both cases, comparison with bounds coming from colliders are shown explicitly. The magnetic field normalisation has been set to $10~\mu$G. Astrophysical setup: MED diffusion model, NFW profile with $\rho_\odot=0.43$ GeV/cm$^3$. Purple lines represent present (solid line) and 10-years projection (dot-dashed line) limits coming from Dwarf galaxies. {Brown line in the top panel represents current DM limits based on the measurement of CMB anisotropies (solid) and near future reach based on the Planck data (dashed line).}}
\label{fig:svVSmX1}}
\end{figure} 

\begin{figure}[ht]
\includegraphics[width=0.5\textwidth,angle=0]{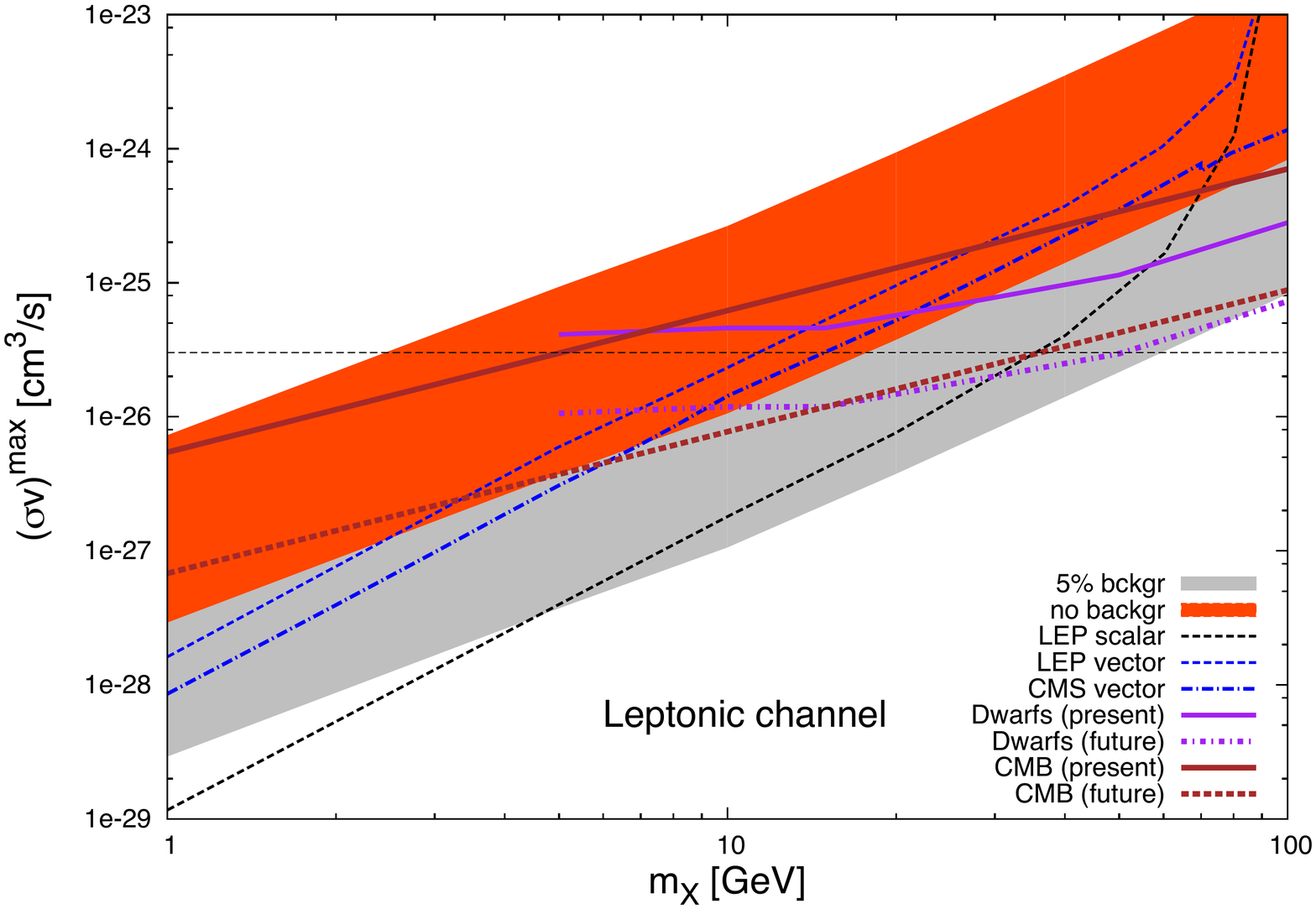}
\includegraphics[width=0.5\textwidth,angle=0]{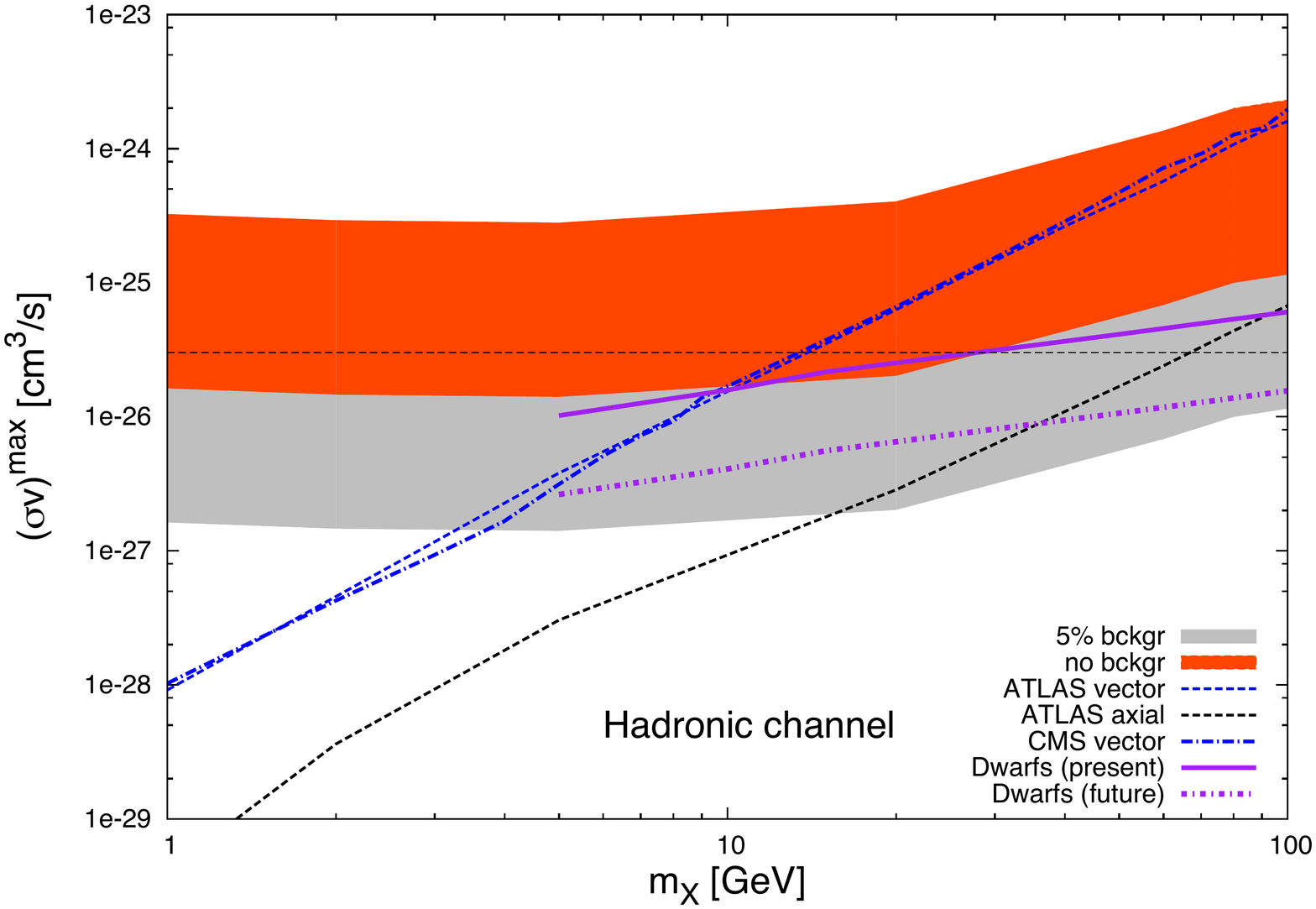}
\caption{\footnotesize{Same as in figure (\ref{fig:svVSmX1}), with the same astrophysical setup, but now the magnetic field normalisation 
has been set to $26~\mu$G.  }}
\label{fig:svVSmX2}
\end{figure} 

\noindent
We also translated these limits on $\Lambda_i$ into limits on  annihilation cross sections.
We present the result in Figures~\ref{fig:svVSmX1} and \ref{fig:svVSmX2} for different values of the magnetic fields, and natures of
the DM couplings, in comparison with limits coming from LEP, CMS, observation of dwarf galaxies by {\it Fermi} LAT and the measurement of CMB anisotropies by the WMAP and ACT.
Comparing the 2 figures, we first notice that the results are not extremely sensitive to the value of the magnetic field (10 or 26~$\mu G$)
and that synchrotron bounds largely compete with bounds from other type of experiments.
Secondly, the hadronic channel gives the weaker bounds on $\langle \sigma v \rangle$, especially at low masses. For low masses, the annihilation into hadronic states gives an electron spectrum much softer,
especially after the kinematic closure of the $b \bar b $ channel, and 
radiates much less synchrotron emission than in the case of a pure leptonic channel: a lot of soft electron produced by later decays of final state hadrons
will radiate at frequencies lower than 45 MHz and will not be bounded by current radio data. At the high mass end, the bounds coming from indirect detections searches (based on the Galactic synchrotron emission,  dwarf Galaxies or CMB) get weaker because the flux of products of DM self-annihilation is proportional to $1/m^2_{\chi}$. 
We should also emphasise that even the $very$ $conservative$ result, where we supposed that all the  data are generated by DM synchrotron, 
gives already limits competitive with LEP bounds, whereas the 5\% uncertainty bounds could give the best limit obtained by
an indirect detection experiment.



\noindent
Concerning the accelerator constraints, from the study of the $\langle \sigma v \rangle$ predictions for different kind of effective interactions (cf. Appendix \ref{AppA}), one realizes by direct inspection of those expressions that normally the vector operator gives larger values, with respect to the scalar and axial case. This is because its Lorentz structure allows for  $\langle \sigma v \rangle$ to develop a term independent on $v^2$ and proportional to $m^2_\chi$, and thus not suppressed. The consequence of this is that in general, vector interaction produce weaker bounds.

\noindent
In the comparison with accelerator constraints, one realises that in the case of leptonic channel, the synchrotron 
bounds in the conservative case of assuming no background, (orange band) is in general as competitive as the bound coming from 
LEP studies, assuming a vector effective interaction. It starts to be more competitive for masses larger than tens of GeV, and beyond 
100 GeV, space of parameter space  beyond the reach of the accelerator. In the case of CMS bound from a vector interaction, 
it is all the way slightly better (for light masses) or similar (for heavier masses) than synchrotron constraints, obviously up to the TeV range 
where LHC starts loosing sensitivity. The LEP bound coming from a scalar effective operator is however one order of magnitude stronger  for
 light masses of $\simeq$ 10 GeV.  Furthermore, to have an idea of how well we would need to understand the background in order to render
  synchrotron more competitive than LEP or CMS, we include in Figure~\ref{fig:svVSmX1} a bound assuming that background is known 
  within  5\%, at 95\% CL. We observe that for these already small uncertainties the synchrotron searches start to be more competitive
   (in general) than collider bounds, independently of the DM mass (in the case of LEP).  

\noindent
We also show in Figure~\ref{fig:svVSmX1} the bounds on hadronic channels, now compared to collider bounds on vectorial and axial interactions. 
Again, for the conservative case of no background, synchrotron bounds start to be more competitive than collider bounds for $m_{DM}$ of about tens
 of GeV, in the case of vectorial interaction, for ATLAS as well as for CMS. In the case of axial coupling, collider bounds are
  always stronger. In the optimistic case of allowing at most 5\% of uncertainties in the background, the bounds from synchrotron can exclude those from
   colliders already at 10 GeV in the case of vectorial  interaction. The difference in the behaviour for masses lighter than 5 GeV, 
   with respect to the leptonic case, is due to quark masses thresholds. In the hadronic case, when the $b\bar b$ channel opens,
    it gives an important contribution to the flux, causing the maximum allowed $\sigma v$ value to decrease. Since there no more threshold after the bottom mass (the top is heavier than the ranges considered here), the flux decreases smoothly when $m_{DM}$ increases, 
     as in the  the case of leptonic channels (for which  all thresholds are below $\sim 2$ GeV).    
 
 \noindent
We run a similar analysis in  Figure~\ref{fig:svVSmX2}, with the ``optimum" choice of magnetic field,  $B_{\mathrm{GC}} = 26~\mu$G, studied
 in the previous section. We see in this case an improvement of synchrotron bounds with respect to collider bounds. 
 If the DM interaction is vectorial, the synchrotron can exclude the collider bounds for all the mass range, for the case of leptonic channels, even in the conservative background assumption. For hadronic channel the exclusion is already effective for masses beyond 10 GeV. As before, depending on our knowledge of the background, synchrotron could be able to exclude current collider bounds on axial interaction. 
 
 
 \noindent
To summarise, our analysis shows that synchrotron bounds, beside being complementary, could be potentially stronger than those obtained from collider searches,
 and that extensive studies of the astrophysical background at these frequencies is well motivated. To this purpose consistent 
 progress is expected to be  achieved in the next years 
with the new high quality data coming from the PLANCK mission\footnote{http://www.rssd.esa.int/index.php?project=Planck} and from low frequency arrays like LOFAR\footnote{http://www.lofar.org/}, which will survey low frequencies 10-240 MHz and is the pathfinder for the future SKA facility\footnote{http://www.skatelescope.org/}. It is also
worth noticing that the analysis
 of the Galactic diffuse emission of the {\it Fermi} LAT will bring further insights on  CR production and propagation processes, 
 additionally constraining the Galactic synchrotron emission (for a current status see, \cite{FermiLAT:2012aa,Strong:2011wd}). We leave such dedicated studies for future works. 

\begin{figure}[ht]
\includegraphics[width=0.5\textwidth,angle=0]{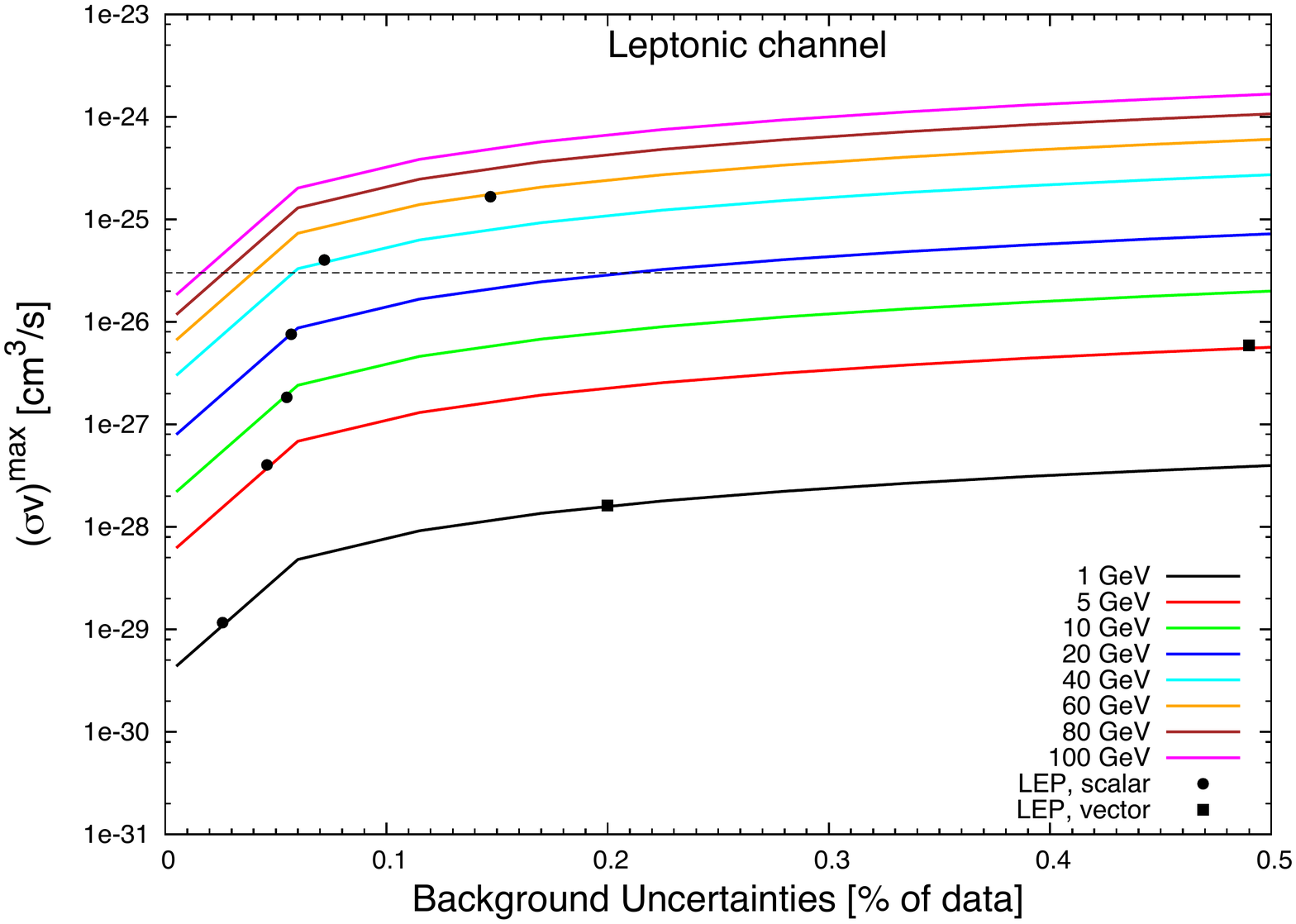}
\includegraphics[width=0.5\textwidth,angle=0]{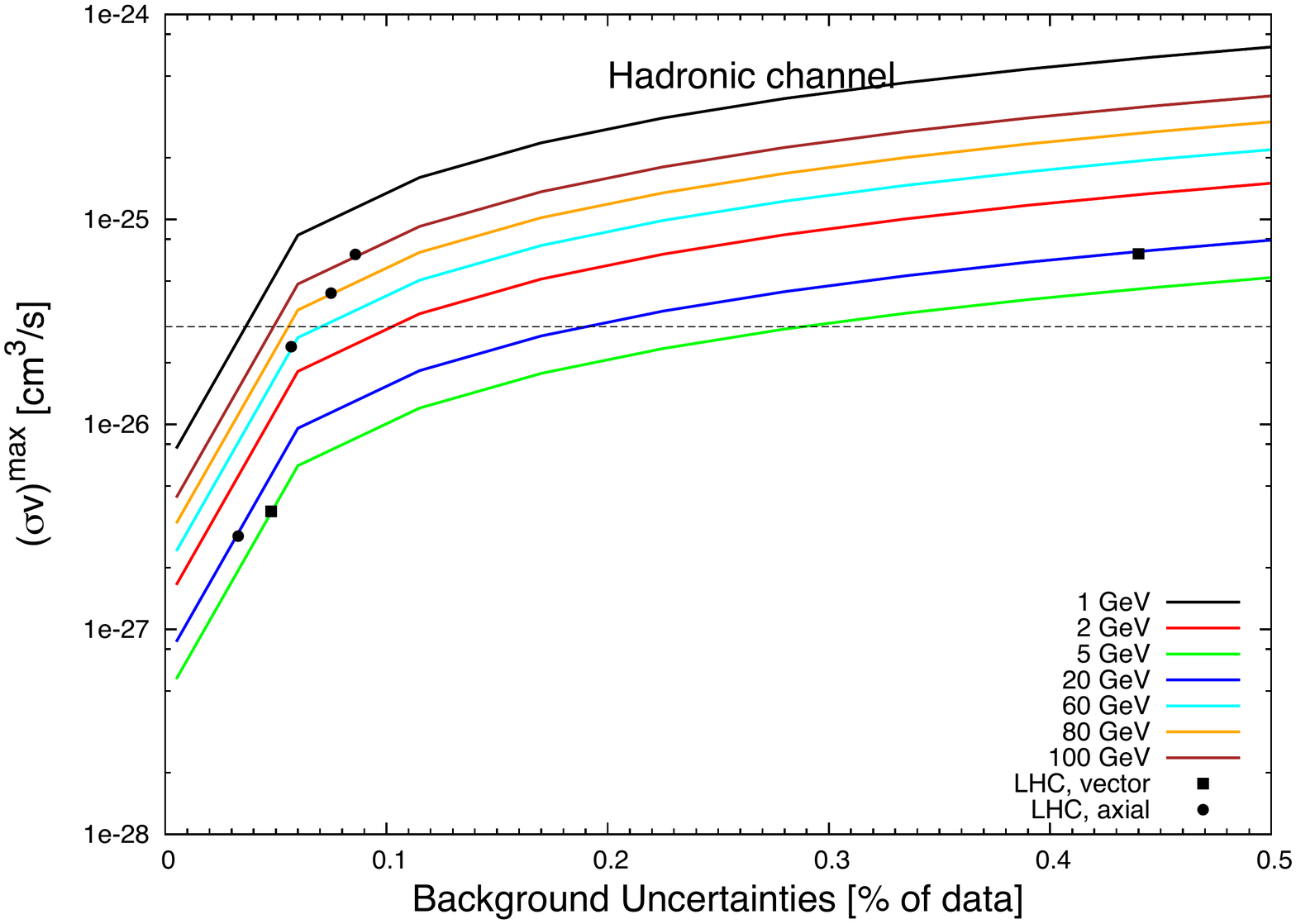}
\caption{\footnotesize{ Bounds on $\langle\sigma v\rangle$ coming from DM synchrotron fluxes as function of background uncertainties assumptions [in \% of data] at 95\%CL. Different choices for DM mass are plotted. Dots represent the bounds coming from LEP (left-panel), specifically the case of a democratic coupling with leptons (as the synchrotron bounds) and assuming a scalar interaction; or LHC-ATLAS (right-panel), for democratic coupling with quarks, for both vector and axial interactions. The magnetic field normalisation has been set to $10~\mu$G. Astrophysical setup: MED diffusion model, NFW profile with $\rho_\odot=0.43$ GeV/cm$^3$.}}
\label{fig:backgr1}
\end{figure} 
\begin{figure}[ht]
\includegraphics[width=0.5\textwidth,angle=0]{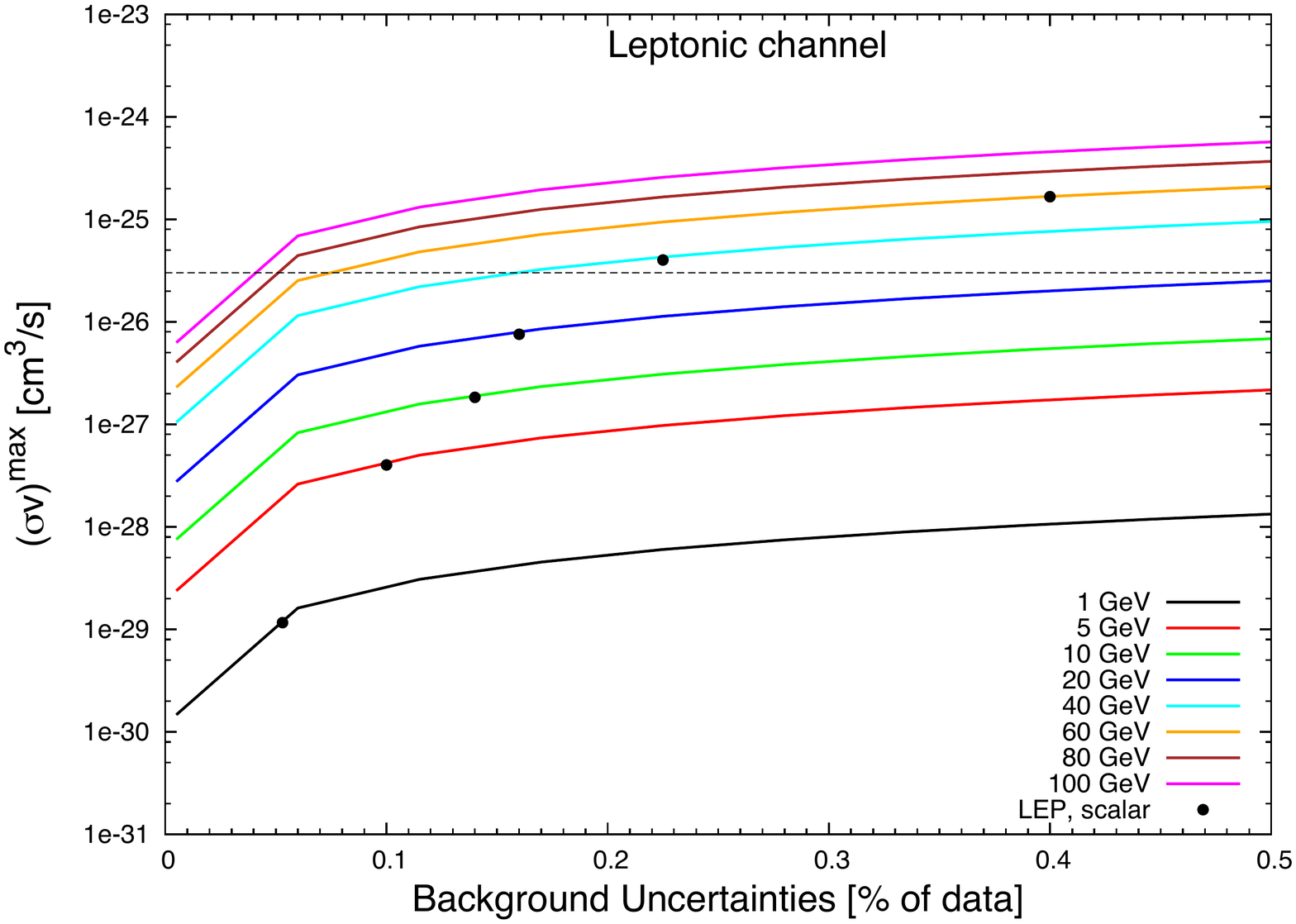}
\includegraphics[width=0.5\textwidth,angle=0]{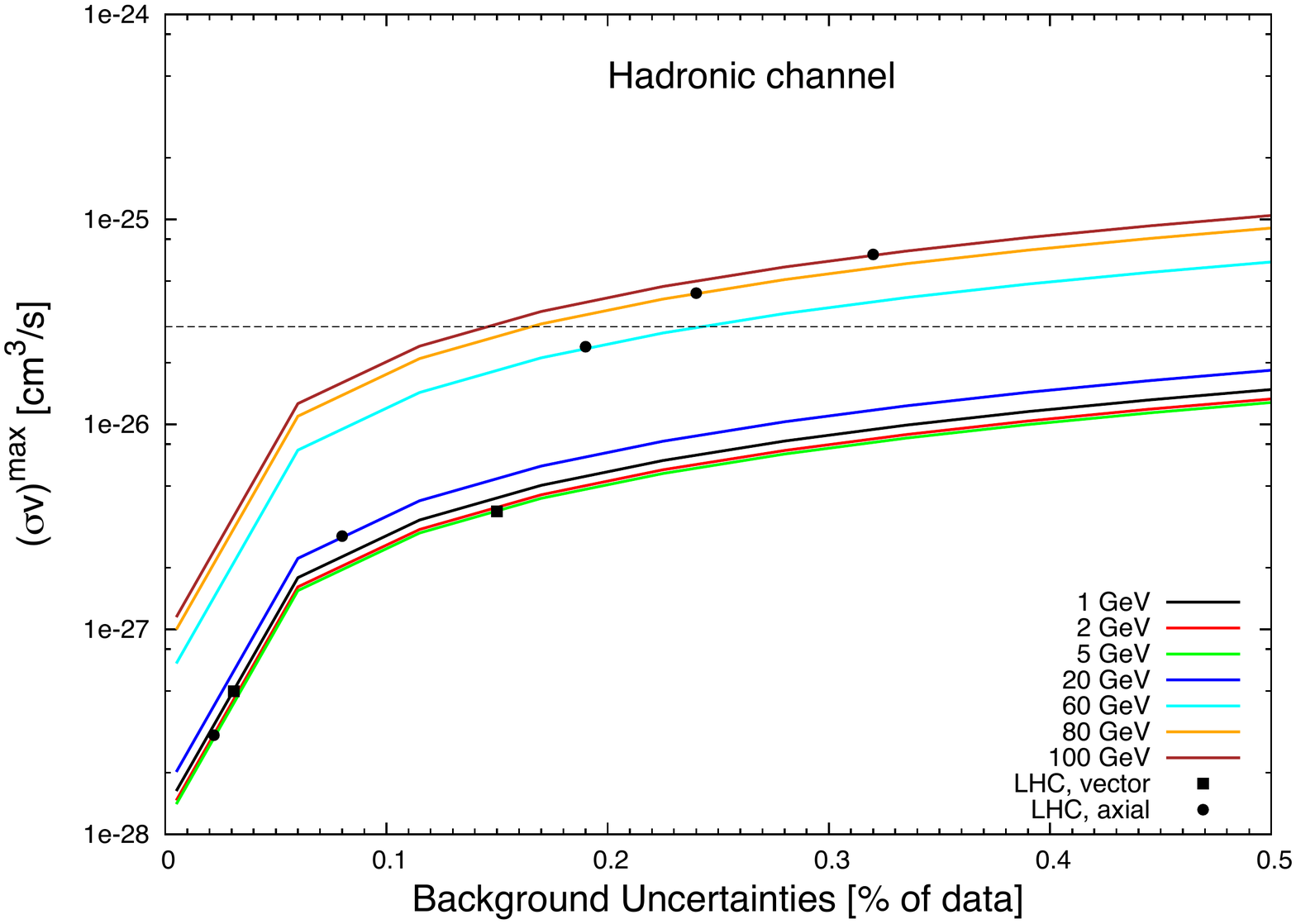}
\caption{\footnotesize{ The same as in Figure~\ref{fig:backgr1}, with the same astrophysical setup, but with the magnetic field normalisation set to $26~\mu$G.}}
\label{fig:backgr2}
\end{figure} 

\noindent
The impact of the control of the background is  illustrated in Figs. \ref{fig:backgr1} and \ref{fig:backgr2}
where we show the level of precision that would be required to be able to exclude/probe the effective models for different DM masses.
The dots on top of the iso--mass lines represent collider bounds for the corresponding mass. The lines with no dots 
 correspond to cases for which  the synchrotron limit is  (for the whole range of uncertainties) stronger or  weaker than the respective collider
 bounds. So, for each iso--mass, the regions on the left-hand-side of the dots correspond to situations for which the synchrotron bounds are stronger, and on the right-hand-sides, to stronger bounds from  colliders.
For instance,  the case of a 20 GeV DM candidate with scalar leptonic interactions is more constrained by radio synchrotron data, provided uncertainties on the background are smaller than 5\%, than by the LEP data (Figure~\ref{fig:backgr1} -- top). For vector-like interactions, the synchrotron 
bounds were always stronger, no matter what the uncertainties considered, for DM masses above 5 GeV and assuming $B_{\mathrm{GC}} = 10~\mu$G 
(see Figure~\ref{fig:backgr1}). For $B_{\mathrm{GC}}=26~\mu$G, (Figure~\ref{fig:backgr2} -- top), the synchrotron bounds are always better than collider bounds, in the case of vector interaction.
 The conclusion is that, for uncertainties on the background of 5\%, or less, the synchrotron bounds are stronger than the LEP bounds, independently
  of the DM mass and of the nature of the effective DM coupling to SM leptons. 

\noindent
Looking at hadronic channels, we see that for uncertainties of $5-10\%$ (Figure~\ref{fig:backgr1} -- bottom), the synchrotron bounds are
 stronger than those coming from LHC, if the interaction is axial-like and for $B_{\mathrm{GC}}=10~\mu$G. The case of $B_{\mathrm{GC}}=26~\mu$G (Figure~\ref{fig:backgr2} -- bottom)
  is of course even better, and for $m_{\chi}\simeq 60$ GeV, even with background uncertainties of 20\%, the synchrotron can rule out the collider bounds. 
  If the interaction is vector-like, the bounds from  colliders are weaker, as we discussed above, and for masses of 20 GeV and beyond, synchrotron 
  bounds are always stronger.

\subsection{Higgs portal}
\label{sec:HiggsPortal}

The effective operator approach is quite powerful, as it encompasses many possible underlying microscopic theories with a minimum set of free parameters, but it has its limitations. This is in particular the case when other degrees of freedom become relevant at the energies considered (be in the early universe or at colliders). Also, a microscopic theory in general predicts some specific relations between the various effective couplings to which it reduces at low energies. Finally, a specific theory may have implications or predictions, that are hidden in an effective operator approach. For all these reasons, it is of interest to consider more fundamental models of dark matter. Hence the purpose of this section, which is to take our analysis of constraints from synchrotron radiation to a more microscopic level. 

Generally speaking, any fundamental theory
that addresses the
abundance of WIMPs requires two fundamental
ingredients:
\begin{itemize}
\item{A candidate. It should be a massive and weakly interacting particle,
stabilised by a (discrete or continuous, possibly gauged) symmetry.}
\item{A mediator. This component can be an existing particle or a
companion of the dark matter
candidate naturally present by construction. For instance it could be a particle
 (the Higgs, or a new scalar) or a new gauge boson (Z')}. 
\end{itemize}

\noindent
In this and the following section, we study the two simplest extensions of the SM with DM particles, the so-called
Higgs-portal and $Z'$, or kinetic mixing, portal.  In particular we determine their radio synchrotron emission, and combined with the WMAP constraint, we analyse the parameter space that is experimentally allowed.

\noindent
The perhaps simplest extension of the SM consists in the addition of a real singlet scalar
field, that couples to the SM through the Higgs field.
In this case, the mediator is the Higgs boson itself, and the model is
usually called the ``Higgs portal".
Although it is logically possible to generalise this scenario to more than one
singlet, the simplest
case of one singlet already provides a very useful framework to
analyse
the generic implications of an augmented scalar sector.
The most general renormalisable potential involving the SM Higgs doublet
$H$ and the
singlet $S$ is
\bea
{\cal L}&\supset& 
- \frac{\mu_S^2}{2} S^2 -\frac{\lambda_{S}}{4} S^4  -
\frac{\lambda_{HS}}{4}S^2 H^{\dag}H
\nonumber
\\
&-&\frac{\kappa_1}{2} H^\dag H S - \frac{\kappa_3}{3}S^3
\label{Eq:Lagrangian}
\eea
 \noindent
 For the case of interest,  the $S$ particle must be stable to be a dark matter candidate. This is achieved by  imposing a $Z_2$ symmetry on the model, $S\rightarrow -S$, with $H$ unchanged, thereby eliminating the $\kappa_1$ and $\kappa_3$  terms.
 We also require that the true vacuum of the theory satisfies $\langle S
\rangle=0$, thereby precluding
 mixing of $S$ and the SM Higgs boson (and the existence of cosmologically
problematic domain walls). In
this case, the mass of the $S$ is simply 
\be
 m_S^2= \mu_S^2 +
\frac{\lambda_{HS}}{4}v^2
 \ee
with $v=246$ GeV is the vev of the Higgs field, 
 and the H-SS coupling is
 \be
{\cal C}_{HSS} =-\frac{\lambda_{HS}M_W}{2 g}~,
 \ee
 where $M_W$ is the W-boson mass and $g$ the $SU(2)$ gauge coupling.

\noindent
 Different aspects of scalar singlet extension of the SM has already been
studied in
\cite{Scalarmodel}
whereas a  preliminary analysis of its dark matter consequences can be
found in
\cite{Barger:2007im}. Some authors considered the possibility of explain the DAMA and/or
COGENT
excesses \cite{Scalardirect},
whereas  several authors looked at the consequences of earlier XENON100 data
\cite{Scalardirect}.
Other authors looked at the consequences on the invisible Higgs width at
LHC \cite{Invisible} or
the restriction of the parameter space due to an hypothetical 125 GeV Higgs
signal \cite{Higgs125}.
Other studies probed the model by indirect searches
\cite{Scalarindirectgamma, Scalarindirectpositron,Arinaindirect} of $\gamma-$ray
or positrons, but there is not yet and analysis of constraints from synchrotron radiation.

\begin{figure}
\includegraphics[width=0.5\textwidth]{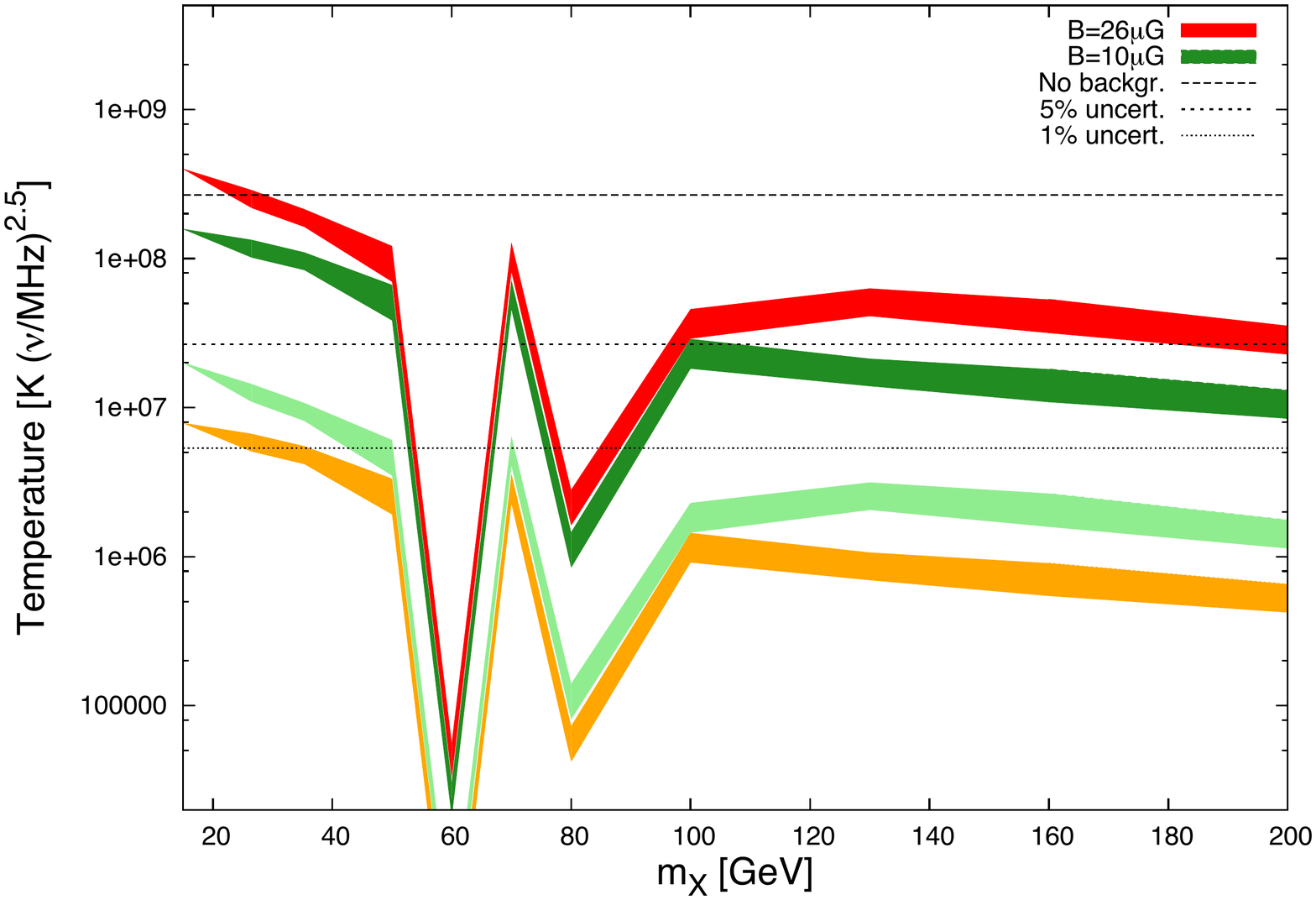}
\includegraphics[width=0.5\textwidth]{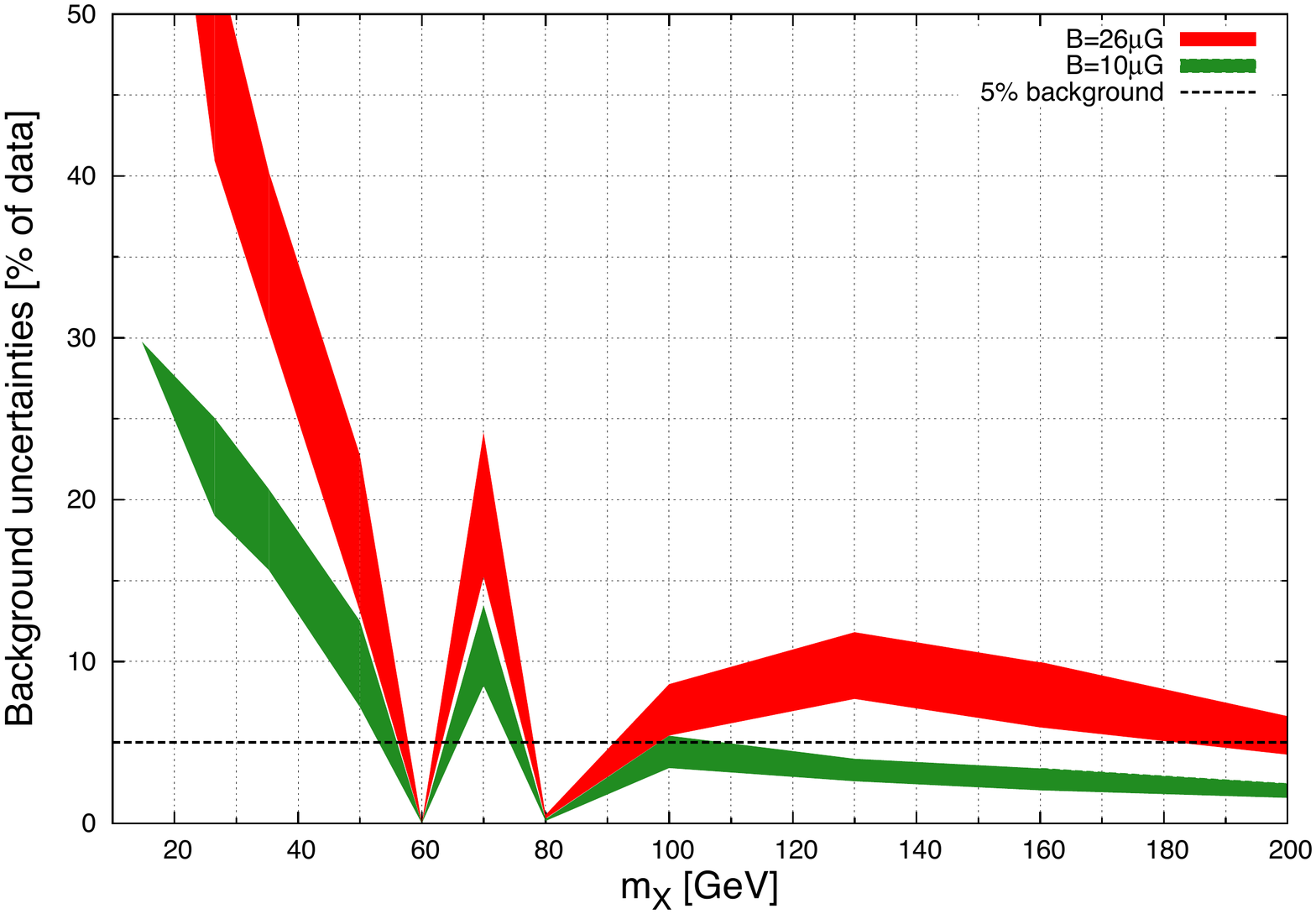}
\caption{\footnotesize{Top) Synchrotron
flux at 45 MHz produced from the Higgs-portal model. For every
$m_{\mathrm{DM}}$, the coupling $\lambda_{\mathrm{HSS}}$ is scanned while
requiring that the relic abundance respects WMAP constraint at 5$\sigma$. 
We presented the result for a
magnetic field profile normalised to 26$\mu$G (red, orange) and 10$\mu$G (green, light green).
Dashed line shows bound coming from data, assuming no background; whereas
dotted line shows the bound assuming a full background with uncertainties
of 5\% (see text for details). {Red and green bands are for NFW+MED+$\rho_\odot=0.43$ GeV\,cm$^{-3}$ astrophysical setup, 
while light-green and orange correspond to ISO+1a model+$\rho_\odot=0.3$ GeV\,cm$^{-3}$.}
 \noindent
 Bottom) Maximum background uncertainties allowed to
exclude a point in the parameter space of the model, with the same conventions as above. }}
\label{Fig:Higgs}
\end{figure}

\noindent
We show in Figure~\ref{Fig:Higgs} (top) the synchrotron flux in our ROI expected for the
Higgs-portal model
for two different representative values of the magnetic fields (10 $\mu$G and
26 $\mu$G)
as a function of the dark matter mass,  for a Higgs of 125 GeV and candidates
that respect the WMAP constraint
(at 5$\sigma$). We also represented the limits we obtain based on the 45 MHz radio data in the two cases discussed above: when we supposed that all the
signal is due
to synchrotron radiation (labelled "no backgr", extremely conservative
case A), and
when we allowed the signal to lie within the 5\% of uncertainties due to
modelling of an astrophysical background of standard origin
(labelled "5\% background, less conservative case B). We do this for two typical astrophysical setups, as studied above in the paper: a NFW dark matter profile, with MED diffusion model and a local normalisation of a DM density of $\rho _{\odot}=0.43$ GeVcm$^{-3}$, and on the other hand, an Isothermal profile, with diffusion model "1a" \footnote{Those models, where Alfv\'en velocity can not be neglected -- at least for our energies of interest-, can not be studied directly by the Semi-Analytical approach followed here. Instead, we use the full numerical analysis to estimate suppression factors in every annihilation channel, and from there we derive the new bounds.} (see Table I) and a local normalisation of a DM density of $\rho _{\odot}=0.3$ GeVcm$^{-3}$. 

\noindent For the (NFW + MED) setup, we observe that, in the more conservative case, dark matter masses below
$\simeq 20$ GeV are already
{close to be} excluded for small values of the magnetic fields ($B = 10~\mu$G),
and even DM masses
below {30} GeV are excluded for $B = 26~\mu$G.
This result is better that the one can obtain based on the {\it Fermi} LAT gamma--ray and PAMELA positron data
\cite{Scalarindirectgamma,Scalarindirectpositron},
which is one of the important results of our analysis.
Indeed, dwarf galaxies study exclude only $m_{s} \lesssim 10$ GeV
\cite{Arinaindirect}.
Also, with the same assumption on the CR propagation model (MED), one can see
that the current synchrotron constraints already reach a higher level of precision
than those based on the PAMELA positron data (see Fig.~2 of \cite{Scalarindirectpositron})
and gives even better result than the projected AMS--02 sensitivity
(see Fig.~3 of \cite{Scalarindirectpositron}) if the DM signal lies in the 10\% uncertainty of
the background. The exclusion by synchrotron emission is also similar to the one sets by the {\it Fermi} LAT
data on galactic gamma--rays \cite{Scalarindirectgamma}.

\noindent
In  case B, allowing the DM signal to stay within 5\%
uncertainty, we observe
that  mainly all the parameter space of the model is excluded {for masses lighter than 100 GeV}, except in
two narrow regions, namely near the Higgs
pole ($2 m_s \simeq m_H$), and just above the $W^\pm$ threshold
($m_S \simeq 80$ GeV).
Indeed, in a region around the pole,
 the enhancement of the cross section due to the Breit-Wigner
form of the amplitude implies that a very low value of ${\cal C}_{HSS}$ is required to
match the WMAP constraint, $\langle \sigma v \rangle \sim 3 \times 10^{-26}~\mrm{cm^3 s^{-1}}$, as
the amplitude is
proportional to 
$${\cal M}\propto \frac{{\cal C}_{HSS}}{E^2_{CM}-m_H^2 +i
\Gamma_H m_H}$$
with $E_{CM}$ being the CM energy of the annihilating $S$ pair.
However, the synchrotron radiation produced in the Galaxy
comes from $S$ essentially at rest,
whereas its kinetic energy at freeze-out was about
$T\simeq m/25$. Consequently, for annihilation in the Galaxy, the points respecting WMAP are 
 away from the pole and the enhancement is not sufficient
 to counterbalance the small value of ${\cal C}_{HSS}$: the annihilation cross section is
 $\langle \sigma v \rangle _{v\simeq 200 \mrm{km s^{-1}}} \simeq
 10^{-29} \mrm{cm^3 s^{-1}}$. The synchrotron flux is thus largely
reduced in this region
 of the parameter space, explaining the first dip seen in Fig.~\ref{Fig:Higgs}.

 \noindent
 The second dip seen  in Fig.~\ref{Fig:Higgs} is also due for kinetic reasons.
 Indeed, as soon as the $W^\pm$ channel is open, the $W^\pm$ final state
is by far the dominant one
 due to the large gauge coupling of the $W$'s to the Higgs.
 Thus, for singlet candidate with a mass slightly below $M_{W^\pm}$, its kinetic energy
at the freeze-out temperature
 allows its annihilation into $W^{\pm}$, even for small values of ${\cal
C}_{HSS}$, as the process is
 dominated by the gauge interaction and not the Yukawa ones (the $b \bar b$
final state is
 the dominant one for $m_s \lesssim M_{W{\pm}}$). However, nowadays, such
slightly lighter
 singlet cannot annihilate into $W^{\pm}$, as its kinetic energy
has dropped below
 the threshold. Consequently, its main decay in the Galaxy is into $b \bar b$
pairs, but
 with a small value of ${\cal C}_{HSS}$, again giving a very low
synchrotron emission.
 These two narrow regions are thus very difficult to observe with this
type of signal.
 We should perhaps emphasize that the arguments we gave above are valid
for any kind of,
 indirect detection prospect ($\gamma$ or antimatter) and is not specific to
 synchrotron.

 \noindent
 We also show in Fig.~\ref{Fig:Higgs} (bottom)  the
precision on the background  that is required to be able to exclude/discover a synchrotron signal for a
singlet DM through the  Higgs portal. We observe that, except in the two narrow pole
regions discussed above
(where  an unreasonable precision would be needed to measure any type of fluxes), 
if 5--10\% (at 95\% CL) of the observed flux is due to synchrotron radiation, the model is already excluded for $B_{\mathrm{GC}}=26~\mu$G. For $B_{\mathrm{GC}}=10~\mu$G one needs to suppose that $\lesssim 5\%$ of the measurement is due to the annihilation of the singlet to exclude the model. This result is then essentially independent of the precise value of $B_{\mathrm{GC}}$.

\noindent {Of course, these conclusions are weaker for the second astrophysical setup considered here, where electrons diffuse more and the DM is less cusped in the inner galaxy. Independently of the magnetic field used, the model is excluded if background uncertainties are of 1\% (at 95\% CL) or less, for DM masses below 40 GeV. Again, as commented above for the case of the effective approach, an equally valid option for the DM distribution, as can be the Einasto profile (not shown here), can enhance the signal by a factor 3, allowing us to claim stronger conclusions about the exclusion power of the synchrotron radiation. } 

\begin{figure}
\includegraphics[width=0.5\textwidth]{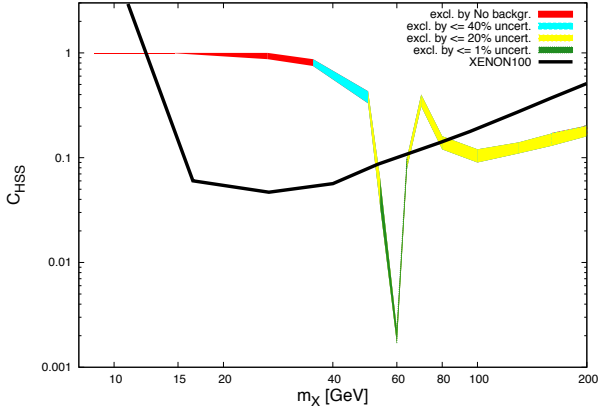}
\caption{\footnotesize{Constraints on the coupling $\cal{C}_{HSS}$ for the Higgs-portal model, as a function of DM mass $m_{\chi}$. Different colours show the exclusion by synchrotron data, assuming: Red) Excluded by No-background choice, Blue) Excluded by uncertainties of 40\%, Yellow) Excluded by uncertainties of 20\%, and Green) Excluded by uncertainties of less than 1\%. Here we assume 26$\mu$G for the magnetic field at GC, as well as NFW+MED set-up. Black solid line shows the exclusion by XENON100. }}
\label{Fig:Lhss}
\end{figure}

\noindent
Finally, some words regarding the comparison with XENON100 bounds. It was shown in \cite{Farina:2011bh} that in general the scalar singlet model is disfavoured by XENON100 data in the region of low ($\lesssim M_W$) masses, except around the region of the Higgs-pole, and also for $m_\chi\lesssim 8$GeV, where XENON100 starts loosing sensitivity.
 In this sense, synchrotron bounds are complementary to XENON100 bounds, as they are able to constrain the region of very low mass, as well as the region of large mass ($\gtrsim M_W$), given some reasonable knowledge about the background.  For illustration, we show in Fig. \ref{Fig:Lhss} the exclusion of the coupling $\cal{C}_{HSS}$ due to synchrotron data, assuming a value of 26$\mu$G for the magnetic field in our ROI, and compare it directly with the exclusion coming from XENON100. We see that effectively for $m_\chi \gtrsim M_W$, synchrotron constraints are stronger than those of XENON100, assuming reasonable backgrounds uncertainties of 20\% at 95\%CL. Also, as expected, synchrotron bounds (even in the conservative case of no-background) do better than XENON100, for masses $m_\chi\lesssim 12$ GeV.
    
\noindent
One should notice that our analysis can also be applied to a vectorial dark
matter through the Higgs portal \cite{Vectordm}.
Indeed, one can consider, to first approximation, a massive vector as a scalar
particle with three internal degrees
of freedom. In this case the synchrotron
flux should be a factor $\simeq 3$ times
greater than
the one we obtained for the simplest singlet scalar extension. One thus
deduces that the constraints
we would obtain would be even stronger than the one we presented in the
singlet scalar scenario.

\subsection{Extra U(1)}
\label{sec:ZpPortal}
Neutral gauge sectors with an additional dark $U'(1)$
symmetry in addition
to the Standard Model (SM) hypercharge $U(1)_Y$ and an associated $Z'$
are among the best motivated extensions of the SM, and give the possibility
that a dark matter candidate lies within this new gauge sector of the theory.
Extra gauge symmetries are predicted in most Grand Unified Theories (GUTs)
and appear systematically in string constructions. Larger groups than $SU(5)$
or $SO(10)$ allow the SM gauge group and $U(1)'$ to be embedded into
bigger GUT groups.
Brane--world $U'(1)$s are special compared to GUT $U'(1)$'s because there
is no reason for the SM particles to be charged under them;
for a review of the phenomenology of the extra $U'(1)$s generated in such
scenarios see e.g.
 \cite{Langacker:2008yv}.
In such framework, the extra $Z'$ gauge boson would act as a portal
between the ``dark world''
(particles not charged under the SM gauge group) and the ``visible'' sector.

\noindent
Several papers considered that the key of the portal could be the gauge
invariant kinetic mixing $(\delta/2) F_Y^{\mu \nu} F'_{\mu \nu}$ \cite{Holdom}.
One of the first models of dark matter from the hidden sector with a
massive additional $U'(1)$, mixing with the SM hypercharge through both mass and kinetic
mixings,
can be found in \cite{Feldman:2006wd}.
The DM candidate $\psi_0$ could be the lightest (and thus stable) particle of
this secluded sector. Such a mixing has been justified in
recent string constructions \cite{Cicoli:2011yh,Kumar:2007zza,
Javier,Cassel:2009pu}, but has also been studied within a model independent
approach \cite{Feldman:2007wj,Pospelov:2008zw,Mambrini:2010yp}.
For typical smoking gun signals in such models, like a monochromatic gamma-ray line, see 
\cite{monogamma}.

The matter content of any dark $U(1)_D$ extension of the SM can be decomposed
into three families of particles:

\begin{itemize}
\item{The $Visible$ $sector$ is made of particles which are charged under the SM
gauge group $SU(3)\times SU(2)\times U_Y(1)$ but not charged under $U_D(1)$
(hence the ``dark'' denomination for this gauge group)}.
\item{the dark sector is composed by the particles charged under
$U_D(1)$ but neutral with respect to the SM gauge symmetries. The dark matter
($\psi_0$) candidate is the lightest particle of the dark sector}.
\item{The $Hybrid$ $sector$ contains states with SM $and$ $U_D(1)$ quantum numbers. These states are fundamental because they act as a portal between
the two previous sectors through the kinetic mixing they induce at loop
order.} 
\end{itemize}

\noindent
From these considerations, it is easy to build the effective Lagrangian
generated at one loop :
\begin{eqnarray}
{\cal L}&=&{\cal L}_{\mrm{SM}}
-\frac{1}{4} \tilde B_{\mu \nu} \tilde B^{\mu \nu}
-\frac{1}{4} \tilde X_{\mu \nu} \tilde X^{\mu \nu}
-\frac{\delta}{2} \tilde B_{\mu \nu} \tilde X^{\mu \nu}
\nonumber
\\
&+&i\sum_i \bar \psi_i \gamma^\mu D_\mu \psi_i
+i\sum_j \bar \Psi_j \gamma^\mu D_\mu \Psi_j\,,
\label{Kinetic}
\end{eqnarray}
$\tilde B_{\mu}$ being the gauge field for the hypercharge, 
$\tilde X_{\mu}$ the gauge field of $U_D(1)$ and
$\psi_i$ the particles from the hidden sector, $\Psi_j$ the particles
 from the hybrid sector, 
$D_{\mu}  =\partial_\mu -i (q_Y \tilde g_Y \tilde B_{\mu} + q_D \tilde g_D
 \tilde X_{\mu} + g T^a W^a_{\mu})$, $T^a$ being the $SU(2)$ generators, and 
\begin{equation}
\delta= \frac{\tilde g_Y \tilde g_D}{16 \pi^2}\sum_j q_Y^j q_D^j 
\log \left( \frac{m_j^2}{M_j^2} \right)
\end{equation}
with $m_j$ and $M_j$ being hybrid mass states \cite{Baumgart:2009tn} .

\noindent
Notice that the sum is on all the hybrid states, as they are the only ones which can contribute to the
 $\tilde B_{\mu},\, \tilde X_{\mu}$ propagator.
After diagonalising of the current eigenstates that makes the gauge kinetic
terms of Eq.~(\ref{Kinetic}) diagonal and canonical, 
we can write after the $SU(2)_L\times U(1)_Y$ breaking\footnote{Our notation
for the gauge fields are 
($\tilde B^\mu,\tilde X^\mu$) before the diagonalization, 
($B^\mu, X^\mu$) after diagonalization and 
($Z^\mu,Z'^\mu$) after the electroweak breaking.}
\begin{eqnarray}
A_{\mu} &=& \sin \theta_W W_{\mu}^3 + \cos \theta_W B_{\mu}
\\
Z_{\mu} &=& \cos \phi ( \cos \theta_W W_{\mu}^3 - \sin \theta_W B_{\mu})
- \sin \phi  X_\mu
\nonumber
\\
Z'_{\mu}&=&\sin \phi (\cos \theta_W W_\mu^3 - \sin \theta_W B_\mu)
+ \cos \phi  X_\mu
\nonumber
\end{eqnarray}
with, to first order in $\delta$,
\begin{eqnarray}
\cos \phi &=& \frac{\alpha}{\sqrt{\alpha^2 + 4 \delta^2 \sin^2 \theta_W}}
~~
\sin \phi = \frac{2 \delta \sin \theta_W}{\sqrt{\alpha^2 + 4 \delta^2 \sin^2 \theta_W}}
\nonumber
\\
\alpha &=& 1- M^2_{Z'}/M^2_Z - \delta^2 \sin^2 \theta_W
\label{sphi}
\\
&\pm& \sqrt{(1-M^2_{Z'}/M^2_Z -\delta^2 \sin^2 \theta_W)^2+ 4 \delta^2 \sin^2 \theta_W}
\nonumber
\end{eqnarray}
\noindent
and + (-) sign if $M_{Z'}< (>)M_Z$.
The kinetic mixing parameter $\delta$ generates an effective coupling of 
SM states $\psi_{\mrm{SM}}$ to $Z'$, and a coupling of $\psi_0$ to 
the SM $Z$ boson which
induces an interaction on nucleons.
Developing the covariant derivative on SM and $\psi_0$ fermions state,
we computed the effective $\psi_{\mrm{SM}}\psi_{\mrm{SM}}Z'$  and
 $\psi_0\psi_0Z$ couplings to first order\footnote{  One can find
  a detailed analysis of the spectrum and couplings of the model
   in the appendix of Ref.\cite{Chun:2010ve}.} in $\delta$ and obtained
   \bea
   {\cal L}= q_D \tilde g_D (\cos \phi~ Z'_\mu \bar \psi_0 \gamma^{\mu} \psi_0 +
   \sin \phi~ Z_\mu \bar \psi_0 \gamma^{\mu} \psi_0 ).
   \eea
 We took $q_D \tilde g_D=3$ trough our analysis, keeping in mind that our results stay 
 completely general by a simple rescaling of the kinetic mixing $\delta$.

\begin{figure}[t]
\includegraphics[width=0.5\textwidth]{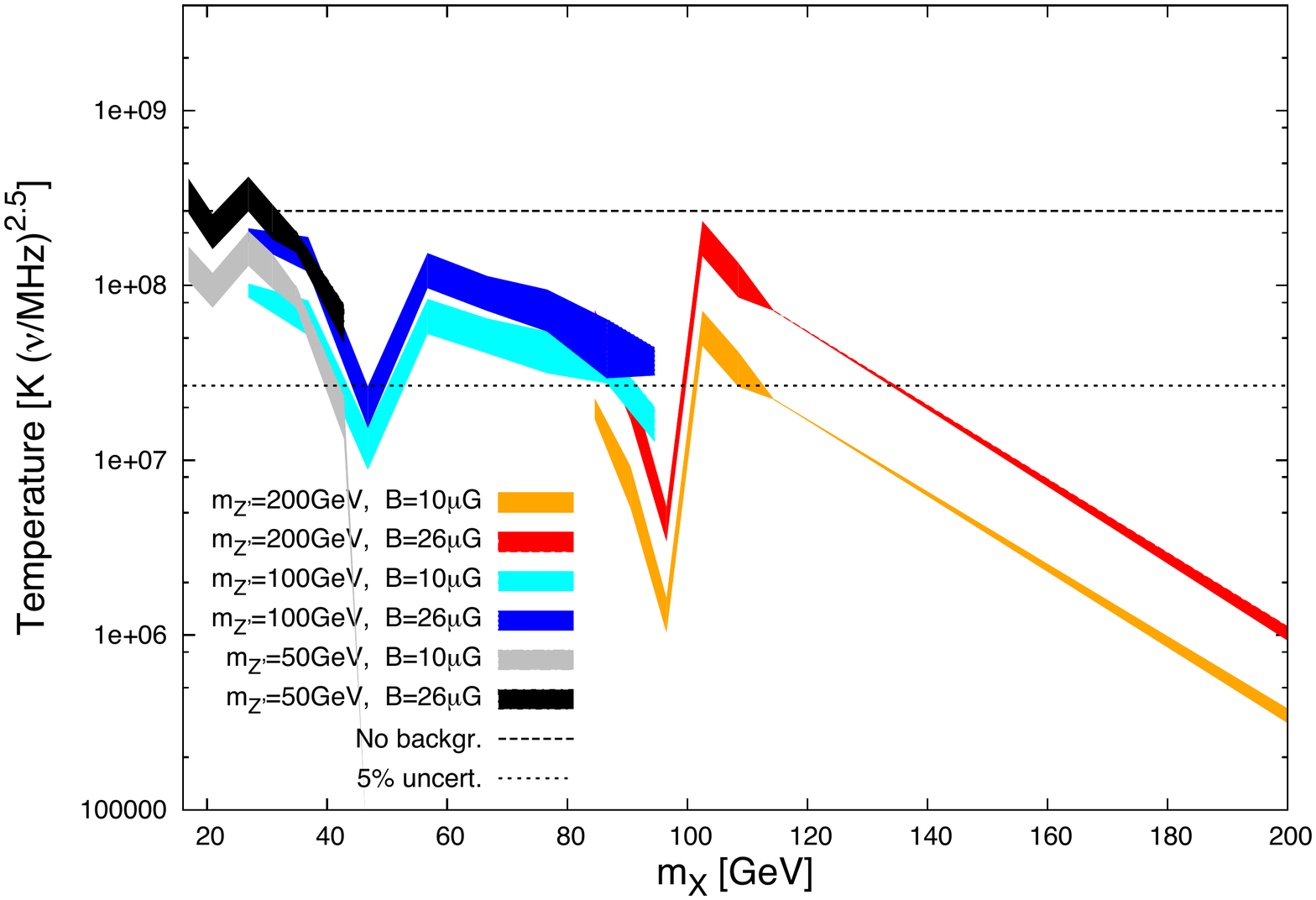}
\includegraphics[width=0.5\textwidth]{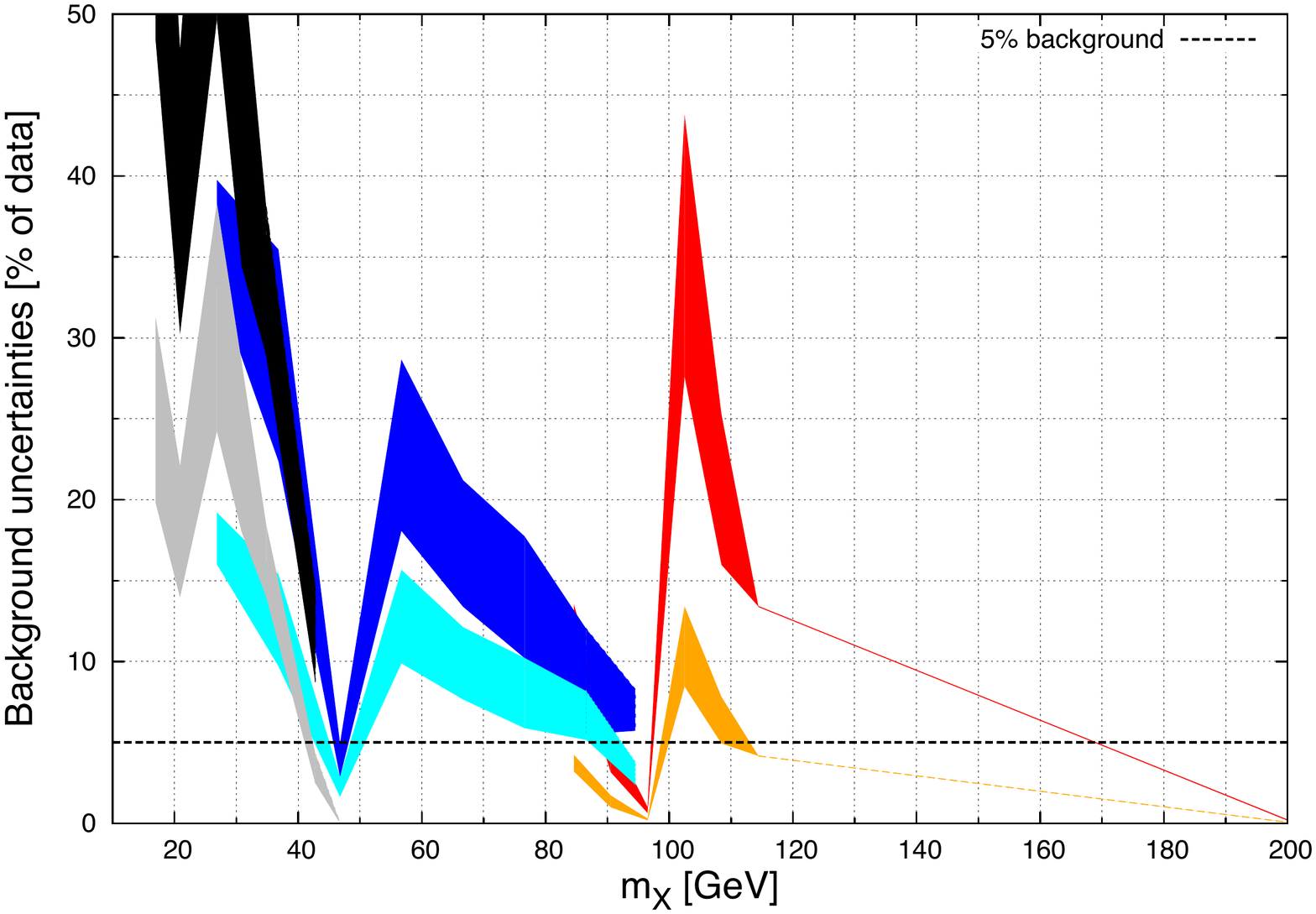}
\caption{\footnotesize{Top) Scatter plot of predictions for synchrotron
flux coming from a $Z'$-portal model defined in (\ref{Kinetic}). For every
$m_{\mathrm{DM}}$, the coupling $\delta$ is scanned while requiring good
relic abundance, for different $Z'$ masses, $m_{Z'}$. A frequency of 45
MHz was used, assuming a magnetic field profile normalised to 26$\mu$G
(dark colours) and 10$\mu$G (light colours). Dashed line shows bound coming
from data, assuming no background; whereas dotted line shows the bound
assuming a full background with uncertainties of 5\% at 95\%CL.  
\noindent
Bottom) Maximum
background uncertainties allowed to be able to exclude a point in the
parameter space of the model. }}
\label{fig:Zp}
\end{figure}

\noindent
We show in Figure~\ref{fig:Zp} the synchrotron flux emitted at 45 MHz for different values of $Z'$ masses after a scan
on ($m_{DM}=m_{\psi_0}; \delta$) for points respecting the WMAP and electroweak precision tests
 (including Z width, $\rho$ parameter $g-2$ and atomic parity violation constraint). {We do this exercise for the astrophysical NFW+MED setup.}
We first observe a similar behaviour to that in the Higgs portal around two poles,  $M_Z$ and $M_{Z'}$.
Indeed, the two channels giving a good relic density are the $s-$channel exchange of the  $Z^{\prime}$ bosons
for $m_{DM}\simeq M_{Z^{\prime}}/2$. We have restricted the fluxes for each $M_{Z'}$ in windows to avoid
juxtapositions of the fluxes. However, when $M_{Z'}\simeq 200$ GeV the value of the flux
begins to be quite weak for regions of other parameter space 
far from the pole region, due to the reduced factor in the amplitude square
of the $s-$channel process $\bar \psi_0 ~\psi_0 \rightarrow Z' \rightarrow SM ~ SM$
$|{\cal M}|^2 \propto 1/M_{Z'}^4$ which is almost a factor 1/6 compared to a 125 GeV Higgs exchange,
factor that we recover in the synchrotron fluxes. In this case one needs a precision better than 5\%
to be able to distinguish a dark matter signal if $m_{DM} \gtrsim 120$ GeV (see Figure~\ref{fig:Zp} -- bottom).

\noindent
One observes that the conclusions we have obtained for the Higgs portal are also quite valid for the $Z'$ portal
too, except that there are  different fine-tuned pole regions for each mass of the $Z'$. The fluxes in both
cases lie in a region of parameter space between {5\%} to 100\%  of the measured flux and could thus
be probed in a near future.

\noindent
\section{Conclusions and prospects}
\label{sec:conclusion}

\noindent In this work, we derived the constraints on synchrotron emission from the secondary products of dark matter annihilation in our Galaxy, based on the radio 45 MHz data. We expanded upon the current literature in several aspects: a) by using both a semi-analytical approach to model the particle propagation in the intergalactic medium \cite{Baltz:1998xv,Delahaye:2007fr}, and a full numerical analysis based on the \texttt{GALPROP} code \cite{Porter:2008ve,Vladimirov:2010aq}, benefiting from the strengths of both in particular aspects of the work, b) considering intermediate Galactic latitudes, where the DM profiles are more robustly constrained,
and c) taking into account the large astrophysical uncertainties on the magnetic fields. We have shown that synchrotron
emissions can give very pertinent bounds on DM annihilation cross section, confirming previous results \cite{Bergstrom:2008ag,Borriello:2008gy,Boehm:2010kg,Fornengo:2011iq}. 

\noindent 
The discussion of the astrophysical propagation setup, 
occupies a large fraction of this work. 
Together with a usual set of CR propagation parameters used previously to gauge uncertainties of the astrophysical conditions on the radio signals, (MIN, MED and MAX models of propagation \cite{Donato:2003xg}), we have considered several other sets of CR parameters, which are shown to be consistent with CR, gamma ray and/or synchrotron data \cite{FermiLAT:2012aa,Trotta:2010mx,Strong:2011wd}. 
We have also explored the impact of the magnitude of the magnetic field in our ROI and found that it is generally mild, ranging a factor of 3--4 for magnetic fields in the inner Galaxy in the range between 1--100~$\mu$G. 

One of the purpose of our work is to put in perspective various indirect bounds on dark matter models, complemented with collider constraints. Although the scope of our results is broader, this was in part motivated by the fact that both colliders and indirect searches are supposed to put very relevant limits on relatively light dark matter candidates, $m_{DM} \lsim 10$ GeV, for which  direct detection limits are weaker, or altogether inexistent. In this spirit, 
we have briefly reviewed (and applied to particle physics models in question, including forecasts) the constraints from {\it Fermi} LAT observations of nearby dwarf spheroidal galaxies \cite{Ackermann:2011wa}, as well as the limits set by CMB anisotropies \cite{Galli:2011rz}. Both of these observations give stringent indirect limits on dark matter annihilation cross sections and treat targets different than the one we focus on to derive the radio DM constraints, providing therefore independent probes of DM self-annihilation signatures. 

We have applied our results to various DM particle models, starting with a generic approach based on effective operators. There, we have shown that, for reasonable values of parameters, and for a conservatively chosen ROI, synchrotron searches for DM could be comparable to those of colliders, and sometimes even do better. In particular, in the case of effective couplings to leptons, we have shown that, even for very conservative setups, where we supposed that all the  radio data are saturated by the synchrotron radiation produced by DM annihilation, the limits obtained are already competitive with those based on LEP measurements (photon+missing energy). In this case too, radio constraints can even be better than the ones derived from CMB, depending on the uncertainty on the astrophysical modelling. Concretely, allowing that DM contribute to  5\% the background signal could give the best limit obtained by current indirect detection signals. 
 For effective couplings to quarks, the synchrotron have a strong potential too, but due to huge uncertainties in the prediction of this signal, one cannot robustly claim that the most conservative synchrotron constraints can improve over the collider bounds.  As expected, if DM couples to hadrons, dwarf constraint generally perform better. 

The effective approach is powerful, but reaches its limits when resonance effects become relevant. To assess such effects, and for the own sake of a more microscopic approach to DM phenomenology, we have considered the constraints set by radio synchrotron radiation on two specific DM models: the so-called Higgs and $Z'$ portals. This is motivated by the fact that these models are among the simplest extensions of the SM with DM candidates. Moreover, each of these models provide fully self-consistent UV completions of the effective operator approach, and thus provide complementary, albeit model dependent, information.  We have shown that radio data may put severe constraints on these models, but at some price. Provided the uncertainty on the background could be assessed at the $5\%$ level, and assuming a cuspy profile (NFW), and a large magnetic field, most of the Higgs portal parameter space is excluded, except near resonance (the Higgs pole), or close the threshold for $W^+W^-$ annihilation, which is impeded in the Galaxy (by construction both these effects are inexistent in an effective operator approach). We have also shown a very good complementarity between synchrotron bounds and the last XENON100 bounds when considering the exclusion of the parameter space of the model. For the case of the $Z^\prime$ portal, because of substantial annihilation into leptons, the constraints are stronger, or alternatively, the astrophysical setup may be more conservative (smaller magnetic field, less cuspy profile), but some control of the background is also required to exclude (most of) the parameter space. 

This work illustrates again, if necessary, that a multi-signal approach provides very complementary information on DM phenomenology. If on the long run, DM production at colliders is likely to give the strongest constraints, one should keep in mind that  missing energy may not be directly related to the actual DM that is supposedly present in our Galaxy. Our results present (part of) the state of art in confronting indirect searches, with a particular emphasis on synchrotron radio data, a very promising signal for dark matter, provided some control may be gained on the mundane, astrophysical background, a fascinating challenge for both future observations and theoretical works. To pave the road, in the near future, PLANCK will be able to study Galactic emission in the frequency range where it is dominated by the dust emission, mapping with unprecedented precision dust (and therefore indirectly gas) content of our Galaxy. Together with improvements in measurement of the charge cosmic ray spectra we will soon be getting from AMS--02\footnote{http://ams.cern.ch/} and measurement of diffuse emission in gamma rays of such CR population, with the {\it Fermi} LAT, models of propagation and energy losses of CR are expected to advance significantly over the next 5-10 year period. Finally, the future radio telescope facilities, like the LOFAR and SKA, will provide further leverage on the possible radio synchrotron signal from DM particles, in particular for lighter candidates.

\section{Acknowledgements}

The authors want to thank Illias Cholis, Alessandro Cuoco, Timur Delahaye, and Roberto Lineros for very useful discussions and opinions. 
Y.M. was supported by
the French ANR TAPDMS {\bf ANR-09-JCJC-0146}  and the Spanish MICINNÕs
Consolider-Ingenio 2010 Programme  under grant  Multi- Dark {\bf
CSD2009-00064}. B.Z. acknowledges the financial support of the  FPI (MICINN) grant BES-2008-004688, and the contracts FPA2010-17747 and PITN-GA-2009-237920 (UNILHC) of the European Commission, as well as LPT-Orsay of Universit\'e Paris Sud, for the hospitality during the development of this project. The work of M.T. is supported by the IISN and an ULB-ARC grant. He also thanks the LPT-Orsay group for support and hospitality. G.Z. is grateful to the Institut d'Astrophysique de Paris for hospitality during completion of this work. 

\newpage
\appendix

\section{Annihilation cross-sections  $\langle\sigma v\rangle$, for different effective operators}
\label{AppA}
The expressions for the annihilation cross-sections coming from the operators $\mathcal O_S, \mathcal O_V$ and $\mathcal O_A$  are given by:

\bea
\sigma_S v &=& \df{1}{8\pi\Lambda^4}\sqrt{1-\df{m_f^2}{m_\chi^2}} (m_\chi^2-m_f^2)v^2~, 
\eea

\bea
\sigma_V v &=& \df{1}{48\pi\Lambda^4}\sqrt{1-\df{m_f^2}{m_\chi^2}}  \\
&\times& \left(24(2m_\chi^2 + m_f^2)+ \df{8m_\chi^4-4m_\chi^2 m_f^2+5m_f^4}{m_\chi^2-m_f^2}v^2\right) ~ , \nonumber
\eea

\bea
\sigma_A v &=& \df{1}{48\pi\Lambda^4}\sqrt{1-\df{m_f^2}{m_\chi^2}} \\
&\times& \left(24 m_f^2+ \df{8m_\chi^4-22m_\chi^2 m_f^2+17m_f^4}{m_\chi^2-m_f^2}v^2\right)  ~;\nonumber
\eea

\noindent
where $m_f$ is the fermion mass and $m_\chi$ the DM mass, while $v$ is the relative velocity of incoming particles, which at the moment of freeze-out it is assumed  to be $v^2=0.24$ \cite{Fox:2011fx}.

\end{document}